\documentclass[usenatbib]{mnras}
\usepackage{hyperref}
\usepackage{color}

\usepackage{graphicx}
\usepackage{amssymb}
\usepackage{pifont}
\usepackage{etoolbox}
\newcommand{\mincir}{\raise
-2.truept\hbox{\rlap{\hbox{$\sim$}}\raise5.truept
\hbox{$<$}\ }}
\newcommand{\magcir}{\raise
-2.truept\hbox{\rlap{\hbox{$\sim$}}\raise5.truept
\hbox{$>$}\ }}
\newcommand{\minmag}{\raise-2.truept\hbox{\rlap{\hbox{$<$}}\raise
6.truept\hbox{$>$}\ }}

\newcommand{\hagn}{\mbox{{\sc \small Horizon-AGN}}}

\newcommand{\hagnn}{\mbox{{\sc \small Horizon-AGN\,\,}}}
\newcommand{\hnoagnn}{\mbox{{\sc \small Horizon-noAGN\,\,}}}

\definecolor{grey}{rgb}{0.75,0.75,0.75}
\definecolor{Orange}{rgb}{1.0,0.5,0.15}
\definecolor{brown}{rgb}{0.7,0.25,0.0}
\definecolor{Pink}{rgb}{1.0,0.5,0.5}
\definecolor{darkerred}{rgb}{0.8,0,0}
\definecolor{darkerblue}{rgb}{0,0,0.8}
\definecolor{Blue}{rgb}{0,0.08,0.65}
\definecolor{Red}{rgb}{0.65,0.08,0.05}
\definecolor{Green}{rgb}{0.15,0.45,0.25}

\def\simlt{\lower.5ex\hbox{$\; \buildrel < \over \sim \;$}}
\def\simgt{\lower.5ex\hbox{$\; \buildrel > \over \sim \;$}}
\def\simpropto{\lower.2ex\hbox{$\; \buildrel \propto \over \sim \;$}}

\title[AGN feedback and total density profile of massive ETGs]
{Total density profile of massive early-type galaxies in Horizon-AGN simulation: impact of AGN feedback and comparison with observations}

\author[S.~Peirani et al.]
 {S\'ebastien Peirani$^{1,2,3,4}$\thanks{E-mail: sebastien.peirani@oca.eu},
 Alessandro Sonnenfeld$^{3}$,
 Rapha\"el Gavazzi$^{2}$,
\newauthor
 Masamune Oguri$^{3,4,5}$,
 Yohan Dubois$^{2}$,
 Joe Silk$^{2,6,8}$,
 Christophe Pichon$^{2,9}$,
\newauthor
Julien Devriendt$^{6,7}$ 
and Sugata Kaviraj$^{10}$\\
$^{1}$ Universit\'e C\^ote d'Azur, Observatoire de la C\^ote d'Azur, CNRS, Laboratoire Lagrange, Bd de l'Observatoire,\\
 CS 34229, 06304 Nice Cedex 4, France \\
$^{2}$ Institut d'Astrophysique de Paris (UMR 7095: CNRS \& UPMC), 98 bis Bd Arago, 75014 Paris, France \\
$^{3}$ Kavli IPMU (WPI), UTIAS, The University of Tokyo, Kashiwa, Chiba 277-8583, Japan \\
$^{4}$ Department of Physics, The University of Tokyo, 7-3-1 Hongo, Bunkyo-ku, Tokyo 113-0033, Japan\\
$^{5}$ Research Center for the Early Universe, School of Science, The University of Tokyo, 7-3-1 Hongo, Bunkyo-ku, Tokyo 113-0033, Japan\\
$^{6}$ Sub-department of Astrophysics, University of Oxford, Keble Road, Oxford OX1 3RH \\
$^{7}$ Observatoire de Lyon, UMR 5574, 9 avenue Charles Andr\'e, Saint Genis Laval 69561, France\\
$^{8}$ Department of Physics and Astronomy, The Johns Hopkins University Homewood Campus, Baltimore, MD 21218, USA\\
$^{9}$ Korea Institute of Advanced Studies (KIAS) 85 Hoegiro, Dongdaemun-gu, Seoul, 02455, Republic of Korea\\
$^{10}$ Centre for Astrophysics Research, University of Hertfordshire, College Lane, Hatfield, Herts, AL10 9AB, UK\\
\\
}

\begin{document}

\maketitle

\begin{abstract}

Using the two large cosmological hydrodynamical simulations,
\hagnn (H$_{\rm AGN}$) and \hnoagnn (H$_{\rm noAGN}$, no AGN feedback),
we investigate how a typical sub-grid model for AGN feedback affects the evolution
 of the total density profiles
 (dark matter + stars)  at the effective radius of massive early-type
 galaxies ($M_* \geq 10^{11} M_\odot$). We have 
studied the dependencies of the mass-weighted density slope $\gamma'_{tot}$  with the effective radius, 
the galaxy mass and the host halo mass  at  $z\sim0.3$ and found that the inclusion of AGN feedback
always leads to a much better agreement with observational values and trends. Our analysis 
suggests also that the inclusion of AGN feedback favours a strong correlation 
between $\gamma'_{tot}$ and the density slope of
the dark matter component while, in the absence of AGN activity,  $\gamma'_{tot}$ is rather strongly correlated with
the density slope of the stellar component.
Finally, we find that $\gamma'_{tot}$   derived from our samples of 
galaxies  increases from $z=2$ to $z=0$,
in good agreement with the expected observational trend. The derived slopes are slightly  lower  than  in the data when AGN
is  included because the simulated galaxies tend to be too extended, especially  the least massive ones. However, the simulated compact galaxies without AGN  feedback  have  $\gamma'_{tot}$ values that are
significantly too  high compared to observations.

\end{abstract}

\begin{keywords}
Galaxies: evolution -- Galaxies: jets -- Galaxies: haloes -- Dark matter --  Methods: numerical
\end{keywords}

\section{Introduction}

Despite the early discovery of tight scaling relations between the physical properties
of the apparently simple ``red and dead'' population of early-type galaxies (ETGs),
the origin and evolution of these relations are still
poorly understood \citep{W+R78,Dre++87,faber87,D+D87}.
These properties concern the stellar mass profile, stellar populations,
metallicity, halo mass to stellar mass relation, as well as stellar mass to
central supermassive black hole mass \citep{magorrianetal98,F+M00,haring&rix04, gueltekinetal09}.
In the standard $\Lambda$CDM paradigm, progenitors of massive collapsed
structures such as massive ETGs form at high redshift by the accretion
of cold filamentary gas \citep{rees&ostriker77, white&rees91, birnboim&dekel03, keresetal05, ocvirketal08, dekeletal09, vandevoort11}
that must be stopped at low redshift by some feedback 
 mechanism associated with 
%mechanics coming from 
the central Active Galactic Nucleus (AGN) \citep{silk&rees98, king03, wyithe&loeb03}.
The emerging picture is that they are formed at a redshift larger
than 3 and almost fully assembled by $z\sim1$ 
\citep[e.g.][]{Ren06}. 
One important conclusion is that their central regions
%part 
must be virialized
 with a self-similar structure as early as $z \sim 1$ \citep{Shet++03}
irrespective of the continuous accretion of dark and luminous matter
 within the $\Lambda$CDM hierarchical merging process. This ETG property
 may be explained if internal regions (within $\sim$ 2 effective radii)
 behave like a dynamical attractor featuring an invariant phase space
 density in spite of their accretion history \citep{loeb03}.
The dark matter and stellar components seem to work together in building
 a ``universal'' nearly isothermal total density profile, even though
 neither of these components is well approximated by an isothermal profile.

In the past few years, the deep study of early-type galaxies playing the role 
of  strong gravitational lenses has led to major developments in the observational side.
Indeed, the strong lensing allows very accurate estimation of the total mass
at the Einstein radius of such systems
and can therefore be combined with  other traditional 
methods such as stellar velocity dispersion \citep{Mir95,N+K96,T+K02a,STE02,New++09},
 and stellar mass maps from multicolor imaging and/or spectroscopy.
Such a combination is  expected to break the degeneracies
inherent to each method alone such as, for instance, the 
 mass-anisotropy degeneracy, bulge-halo degeneracy and the stellar mass/initial mass function (IMF) degeneracy
\citep[e.g.][]{K+T03,T+K04,Tre++10,Aug++10,Dut++13b}.
Weak lensing analysis can also provide additional constraints
though it is not yet possible to use  for individual galaxies \citep{Gav++07,J+K07,Lag++10}.
Another way to break the degeneracy in the analysis of strong lens systems is
to use quasar microlensing which directly measures the stellar mass function 
at the image position \citep[e.g.][]{ORF14,Schechter++14,Jimenez++15}.

%Great observational progress has been achieved in the
%past few years with the systematic study of early-type galaxies acting
%as strong gravitational lenses.
% For these systems, strong lensing
%provides an absolute  measurement of the total mass with a few percent accuracy at a fiducial
%radius (the Einstein radius).
% Used in combination with traditional
%diagnostics such as stellar velocity dispersion \citep{Mir95,N+K96,T+K02a,STE02,New++09},
% and stellar mass maps from multicolor
%imaging and/or spectroscopy, one can break many of the degeneracies
%inherent to each method alone, including the mass-anisotropy
%degeneracy, bulge-halo degeneracy, and the stellar mass/initial mass
%function (IMF) degeneracy \citep[e.g.][]{K+T03,T+K04,Tre++10,Aug++10,Dut++13b}.
%Additional information can be gathered with the addition of
%weak-lensing  \citep{Gav++07,J+K07,Lag++10},
%although at the moment this is not possible for individual galaxies.
%Another way to break the degeneracy in the analysis of strong lens systems is
%to use quasar microlensing which directly measures the stellar mass function 
%at the image position \citep[e.g.][]{ORF14,Schechter++14,Jimenez++15}.

The sample assembled by the SLACS team has shown that lensing galaxies are
indistinguishable from non-lensing early-type galaxies with similar velocity dispersions
in terms of their internal properties and environment
\citep{Bol++06,Koo++06,Tre++06,Gav++07,Bol++08a,Tre++09,Koo++09,Aug++09,Aug++10,Bar++11a,Czo++12,Shu++15}.
Additional samples of higher redshift lenses with measured velocity dispersion have then been built, such
as the Strong Lensing Legacy Survey
 \citep{Tu++09,Gav++14,Ruf++11,PaperIII,Gav++12,PaperIV,Son++15}
or BOSS Emission-line Lens Survey
 \citep{Bro++12,Bol++12,Shu++16a} and
 allowed one
 to investigate the time dependence of the slope of the inner
 total mass density profile near the effective radius \citep[see also][]{Tor++14,Dye++14,Pos++15,Smi++15}.
At fixed mass, the total density slope correlates with the projected stellar mass
 density and   seems to decrease with redshift while remaining close to
 isothermal ($\gamma' \equiv -\mathrm{d} \log \rho / \mathrm{d} \log r \simeq 2$,
 with a noticeably small $\sim 6\%$ intrinsic scatter).
Similar conclusions can be drawn from spatially resolved stellar
 kinematics of ETGs \citep[see e.g.~][for a review]{Cap16}.
In addition, the combination of strong lensing and stellar kinematics
 at group and cluster scales can also provide us with sizable statistical
 samples \citep{New++13b,New++15} for extending our understanding of 
 galaxy formation towards the high mass end.

In parallel to these observational efforts, several attempts were made
 to understand the origin and the tightness of this relation, either with
 semi-analytical models or idealized simulations \citep{Nip++09,L+O10,Nip++12,lackneretal12,hirschmannetal12,JNO++12,Rem++13,dutton2014,SNT14,Sha++17}.
The complex coupling of scales (from dark matter haloes and beyond to star
 formation and AGN activity) involved for  understanding this dynamical
 interplay between baryons and dark matter haloes requires very demanding
 numerical hydrodynamical simulations.
An important step forward was made possible by zoomed hydrodynamical
 simulations around massive galaxies that contain realistic accretion
 scenarios and sufficient resolution to recover with good fidelity the
 internal structure of massive ETGs \citep{Dub++13}.
In particular, a detailed census on the role of the feedback from the
 central AGN was made,  showing that the quenching of star formation
 by the central engine is not only important for  reproducing  stellar
 to halo mass relations and colors but also for reproducing the size
 and the dynamical structure (pressure support instead of rotation).
 The small statistics permitted with these zooms has recently been alleviated
 by  a  hydrodynamical simulation with {\sc ramses} \citep{teyssier2002} of larger cosmological volumes
 containing $\sim 10^5$ objects with similar resolution. This was one of the
 purposes of the \hagnn simulation \citep{Dub++14,Dub++16,Wel++17,kaviraj++17,beckmann++17}, who
 confirmed with better statistics some of the results of \cite{Dub++13}.
%We have recently investigated in \citep{Pei++17} the role of the AGN feedback
% on the evolution of internal dark matter and stellar density profile by 
%using a suite of three large cosmological hydrodynamical simulations, \hagn, \hnoagnn (no AGN
%feedback) and \hdmm (no baryons) with the same initial conditions.

Recently, a comparison of the internal structure of massive ETGs in
 the Illustris Simulation \citep{illustris} which is quite comparable to \hagnn 
 showed reasonable agreement with observations, although the size of
 galaxies seems to be slightly overshot \citep{Xu++16}. This confirms the interest in
 performing detailed comparisons between simulations and observations
 with different simulation codes and solvers in order to study degeneracies
 between subgrid physical recipes and solver specificities that could
 match observations for very different physical reasons. Hence, since
 we already studied in detail the role of AGN feedback 
on the evolution of internal dark matter and stellar density profiles
from our set of \hagnn simulations \citep{Pei++17},
 we will now attempt to compare our results with
 the SLACS+SL2S lensing observational constraints on the internal structure
 of massive galaxies coming from strong lensing and dynamics, the main
 focus being the total mass density profile within the effective radius.

The paper is organised as follows. Section \ref{sec:simu}
briefly the numerical modelling used in this work (simulations and post-processing) while
Section \ref{sec:result} presents our main results relative to the evolution of
 the total density profiles in massive ETGs. 
We finally conclude in Section \ref{sec:disc}.

\section{Numerical modelling}\label{sec:simu}

In this paper, we analyse and compare galaxy samples extracted from 
two large cosmological hydrodynamical simulations, \hagnn 
(H$_{\rm AGN}$) and  \hnoagnn (H$_{\rm noAGN}$). These simulations and galaxies/dark matter haloes samples production
 have been already described in great details in \cite{Dub++14,Dub++16} and \cite{Pei++17}.
In the following, we only summarise the main features.

\subsection{Horizon-AGN and Horizon-noAGN}

%In this paper, we analyse and compare two large cosmological hydrodynamical simulations, \hagnn 
%(H$_{\rm AGN}$) and
% \hnoagnn (H$_{\rm noAGN}$).
%\hagnn is already described in \cite{Dub++14}, so we only summarise here its main features.

 The \hagnn simulation \citep{Dub++14} considers a standard $\Lambda$CDM cosmology with 
total matter density $\Omega_{\rm m}=0.272$, 
dark energy density $\Omega_\Lambda=0.728$, amplitude of the matter power spectrum $\sigma_8=0.81$,
 baryon density $\Omega_{\rm b}=0.045$, Hubble constant $H_0=70.4 \, \rm km\,s^{-1}\,Mpc^{-1}$, and $n_s=0.967$ 
compatible with the WMAP-7 \citep{komatsuetal11}.
The simulation was performed in a periodic box of side $L_{\rm box}=100 \, h^{-1}\rm\,Mpc$
containing  $1024^3$ dark matter (DM) particles, resulting in a DM mass resolution of
 $M_{\rm DM, res}=8.27\times 10^7 \, \rm M_\odot$.
The simulation is run with the {\sc ramses} code \citep{teyssier2002}, and the
 initially uniform grid is adaptively refined down to $\Delta x=1$ proper kpc at all times. 

\hagnn includes gas dynamics, gas cooling and heating (from an uniform UV background
 taking place after redshift $z_{\rm reion} = 10$), and various sub-grid models such as:
 star formation, feedback from stars
 (stellar winds, type Ia and type II supernovae), 
metal enrichment of the interstellar medium by following six chemical species (O, Fe, C, N, Mg, Si). 
Black hole (BH) formation and growth  are also included. 
We consider a  Bondi-Hoyle-Lyttleton accretion rate on to 
 BHs namely
$\dot{M}_{BH}=4\pi\alpha G^2M_{BH}^2 \overline{\rho}/(\overline{c}_s^2+\overline{u}^2)^{3/2}$, 
where $M_{BH}$ is the BH mass,
$\rho$ the mean gas density,  
$\overline{c}_s$ the average sound speed,  
$\overline{u}$  the average gas velocity relative to the BH velocity
 and $\alpha$  a dimensionless boost factor defined by
$\alpha = (\rho/\rho_0)^2$  when $\rho>\rho_0$
and $\alpha= 1$ otherwise \citep{boothandschaye09}.
%in order to account for our inability to capture the colder
%and higher density regions of the interstellar medium.
The effective accretion rate on to BHs is capped at the Eddington accretion rate:
$\dot{M}_{Edd}=4\pi GM_{BH}m_p/(\epsilon_r \sigma_T  c)$
in which 
$c$ is the speed of light,
$m_p$ the proton mass,
$\sigma_T$ the Thompson cross-section and
$\epsilon_r$ the radiative efficiency. In our modelling, we
assumed $\epsilon_r=0.1$ for the \cite{shakuraandsunyaev73}
accretion on to a Schwarzschild BH.
BHs release energy in  two distinct modes:  quasar (heating) 
 or radio (kinetic jet) mode when the accretion rate $\chi \equiv \dot{M}_{BH}/\dot{E}_{dd}$ is
 $\chi > 0.01$ and  $\chi < 0.01$ respectively.
The quasar mode is modelled as an isotropic injection of thermal energy into the gas within a sphere of radius
$\Delta x$ and at an energy deposition rate $\dot{E}_{AGN} = \epsilon_f \epsilon_r \dot{M}_{BH} c^2$,
 where  $\epsilon_f = 0.15$ is a free parameter used  to match the BH-galaxy  scaling relations (see \cite{dubois2012bh} for detail).
At low accretion rates,  energy is released through  AGN radio mode into a bipolar outflow with a jet velocity
 of $10^4 km s^{-1}$. Following \cite{ommaetal04}, a cylinder with a cross-sectional radius $\Delta x$
and height 2$\Delta x$ is used to model the outflow  (see \cite{duboisetal10} for details).
In this case, 
the efficiency of the radio mode is larger than the quasar mode with $\epsilon_f = 1$.
In Fig. 5 of \cite{Pei++17}, we have studied the evolution of
 the Eddington ratio $\chi$ over relevant dark matter halo mass ranges. 
Our results strongly indicate that the radio mode tends to be the dominant mode below $z\lesssim 2$.
These results are also in agreement with \cite{volonteri++16} who have studied in detail
 the cosmic evolution of black holes in the \hagnn simulation.

\hnoagnn was performed using the same set of initial conditions and sub-grid modelling but
 with no black hole formation and therefore no AGN feedback.
 The stellar feedback has not been changed between the two simulations,
and we insist on the fact that the stellar feedback in Horizon-AGN is not ``tuned'' to match any galaxy properties.
In both simulations, the stellar mass resolution is $M_{\rm *,res}=2\times 10^6 \,\rm M_\odot$.

Finally, it is worth mentioning that 
the \hagnn simulation has been extensively analysed in order to make theoretical predictions
to be compared to observational data. In particular, the statistical properties of the simulated galaxies
has been studied, showing good agreement  with observed stellar mass functions all the
way to $z\sim 6$ \citep{kaviraj++17}. The colour and star formation histories are also
well recovered as well as the so called black hole - bulge relations and
duty-cycles of AGNs \citep{volonteri++16}.

\subsection{Galaxy catalogues }

Galaxies are identified using the \mbox{{\sc \small AdaptaHOP}} (sub)halo finder \citep{aubertetal04} and for
the present study, we mainly focus on galaxies with a mass greater than  $10^{11} M_\odot$.
We use also the most bound particle as the definition of their centre.
For each given H$_{\rm AGN}$ or H$_{\rm noAGN}$ galaxy, we compute the radial, tangential and
vertical velocity components of each stellar particle where the orientation  of the  z-axis cylindrical  coordinate
 is defined by the spin vector (i.e. the angular momentum vector from the stellar component).
Then, we estimate the rotational velocity $V$ of the galaxy
by computing the average of the tangential velocity component.
The velocity dispersion $\sigma$ is obtained
from the dispersion of the radial $\sigma_r$, the tangential  $\sigma_\theta$ and the vertical velocity $\sigma_z$
components around their averaged values namely $\sigma^2=(\sigma_r^2 + \sigma_\theta^2 + \sigma_z^2 )/3$. 
In our analysis, we select in general H$_{\rm AGN}$ and H$_{\rm noAGN}$  galaxies with $V/\sigma < 1$ because 
we are interested in ETGs only.

Finally, in order to match galaxies between  H$_{\rm AGN}$ and  H$_{\rm noAGN}$ simulations,
we  use the  scheme developed in \cite{Pei++17}.
To summarize, we first match host dark matter haloes between the two simulations.
To do so, we use the fact that, as we start from the same initial conditions, each dark matter particle
 possesses an identity which is identical among the twin simulations.
 Then, if more than 75\% of the particles of any given halo
 in the \hagnn can also be found in a halo identified in the \hnoagnn, we conclude that these haloes are associated
(provided that their mass ratio is lower than 10 or greater than 0.1).
This scheme cannot be repeated to match galaxies since stellar
particles are created during the simulations and do not necessarily correspond between the two runs.
Therefore, we first couple each galaxy to a 
host dark matter halo in their parent simulation by choosing the most massive galaxy whose centre is
located within a sphere of radius equal to $5\%$ of the virial radius
of its host halo. Galaxy pairs between the two simulation are then determined through
the matching of their host halo as previously described.
Note that past numerical works suggest that the presence of AGN feedback allows mergers to durably 
transform rotationally-supported discs ($V/\sigma > 1$) into dispersion-dominated ellipsoids
($V/\sigma < 1$) \citep{Dub++13,Dub++16}. Therefore, 
when using our matching algorithm,  H$_{\rm AGN}$ galaxies satisfying $V/\sigma < 1$ might be associated to
H$_{\rm noAGN}$ galaxies with  $V/\sigma > 1$.

\section{The total density slopes at the effective radius}\label{sec:result}

\subsection{Definitions}
\label{section_definitions}

In the present study, we focus on the {\it mass-weighted density slope within $r_1$ and $r_2$} 
introduced by \cite{dutton2014}: 

\begin{equation}
\gamma' = \frac{1}{M(r_2)-M(r_1)}\int_{r_1}^{r_2}\gamma(x)4\pi x^2 \rho(x)\, \mathrm{d}x
\end{equation}

\noindent
where $\gamma\equiv - {\it d}$\,log\,${\it \rho}$ $/{\it d}\,$log$\, ${\it r} is the local logarithmic slope
 of the density profile $\rho$ and $M$ the local mass. Using a discrete representation of each density profile,
$\gamma(r)$ and  $M(r)$ can be estimated within each radial bin.

For each studied galaxy, we have considered a
random orientation in space and then derive the effective radius $R_{e}$ at which half of the
projected stellar mass is enclosed. 
%We have also computed the 1d velocity dispersion
%along the line of sight and in a aperture of [0-$R_{e}$/2].
Finally, in order to estimate $\gamma'$ at $R_{e}$, we consider the interval
[$r_1$ - $r_2$]$=$[$R_{e}$/2 - $R_{e}$] since it corresponds to radii probed by the strong lens samples used in this paper.
In strong lensing + dynamics studies, the total density slope $\gamma'$ is observationally determined
 by fitting a power-law density profile to the Einstein radius and central velocity dispersion \citep{Koo++06}.
 Strictly speaking, the value of $\gamma'$ recovered in this way is equal to the mass-weighted slope of the lens
 only if its true density profile is a power-law. Nevertheless, \citet{PaperIV} showed how these two definitions 
of the slope give values of $\gamma'$ that are typically within $0.05$ of each other, for a variety of lens 
density profiles \citep[see also][for a detailed study]{Xu++16}.
Note also that the simulation grid size has a value of 1 kpc where
 dark matter or galaxy density profiles might not fully converge.
In the Appendix A, we have performed a convergence study in the same way than 
 \cite{duffy++10} and found that the lower limit value of $\sim$ 5 kpc
recommended by \cite{power++03} for our studied halo/galaxy mass range, is suitable in our analysis,
 though their work concerns pure dark matter simulations only.
In general, the effective radius of H$_{\rm AGN}$ galaxies is always larger than 5 kpc.
However, this is not the case for  H$_{\rm noAGN}$ galaxies and therefore, in certain cases,
 we will remove galaxies with  too low effective radii (i.e. $R_e/2 \le 1-2$ kpc).
%We estimate $\gamma'$ at the effective radius $R_{e}$ at which half of the
%projected stellar mass is enclosed by considering $r_1=R_{e}/2$ and $r_2=R_{e}$.
%Note that for each galaxy, we compute its effective radius as follow: we compute $R_e$ from
%three different stellar projections $R_{eff1}$ (x-y projection), 
%$R_{eff2}$ (x-z projection) and $R_{eff3}$ (y-z projection).
%Then  $R_e=(R_{eff1}\times R_{eff2}\times R_{eff3})^{1/3}$.

In the following, we will refer respectively to  $\gamma'_{dm}$,  $\gamma'_*$ and  $\gamma'_{tot}$ 
the mass-weighted density slope derived from the dark matter component,
 the stellar component and the dark matter + stellar components.

\subsection{Dependencies of $\gamma'_{tot}$ with $R_{e}$, $M_{halo}$ and $M_*$}

%%%%%%%%%%%%%%%%%%%%%%%%%%%%%%%%%%%%%%%%%%%%%%%%%
%     FIG 1 : log Re vs M*
%%%%%%%%%%%%%%%%%%%%%%%%%%%%%%%%%%%%%%%%%%%%%%%%%
\begin{figure}
\begin{center}
\rotatebox{0}{\includegraphics[width=\columnwidth]{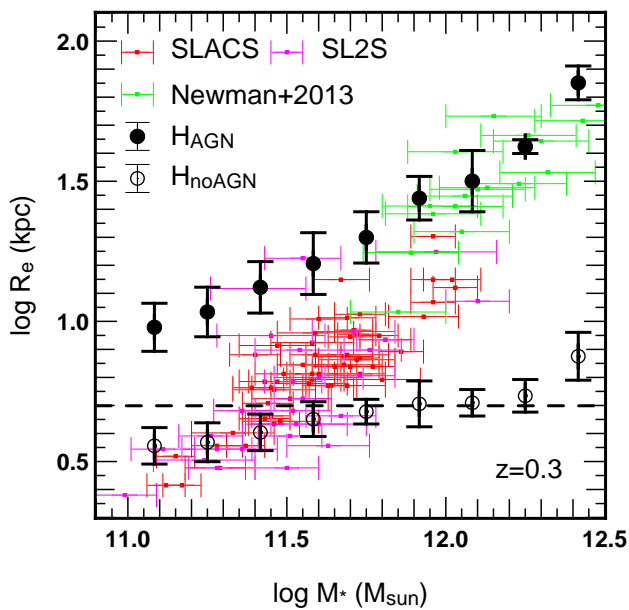}}
\caption{The variations of the effective radius $R_{e}$ with respect to the stellar
mass $M_*$. Results are derived from H$_{\rm AGN}$ (black points) and H$_{\rm noAGN}$ (white points)
 galaxies satisfying $V/\sigma<1$ and  log$(M_*/M_\odot)>11$ at $z=0.3$. We also plot observational data from the SLACS, 
 SL2S and \citet{New++13b} which cover the redshift range of [0.063 - 0.884] but centred around $z\sim 0.3$.
% Note that fitting formula
 %provided in van der Well et al. (2004) have been corrected to consider a Salpeter IMF.
 Error bars indicate the $1\sigma$ standard deviations using a confidence interval on the scatter in a bin.
For indicative purposes only, the horizontal dashed lines at $R_e=$5 kpc
 indicates a recommended resolution limit following \citet{power++03}.}
\label{fig1}
\end{center}
 \end{figure}
%%%%%%%%%%%%%%%%%%%%%%%%%%%%%%%%%%%%%%%%%%%%%%%%%

Fig. \ref{fig1} shows the variations of the effective radius $R_{e}$ with respect to
 the  stellar mass $M_*$  for H$_{\rm AGN}$ and H$_{\rm noAGN}$ galaxies
with a mass greater than $10^{11} M_\odot$ and  $V/\sigma<1$ at $z=0.3$. 
We chose this latter redshift value because the observational data we want to compare with, in particular
SLACS, SL2S and \cite{New++13b,New++15}, are centred around this value.
The observed values of the stellar mass depend upon the assumption of a stellar initial mass
 function (IMF). For our comparison we assume a Salpeter IMF, which is found to be consistent with
 stellar masses derived from the combination of lensing and dynamics in the sample of SLACS and SL2S
 lenses \citep{Son++15}.
First, we see clearly that in the absence of AGN feedback,
the simulated galaxies are generally clearly too compact with an effective radius too small,
 compared to observational data. Note however that for  H$_{\rm noAGN}$ galaxies
 with log$(M_*/M_\odot)<11.5$, the theoretical predictions
seems to  agree well with observations. This might be due to the fact that
as the effective radius get closer to the resolution limit, their value might be overestimated.
On the contrary, AGN feedback tends to form more extended galaxies 
as already noted by \cite{Dub++13}. In this case,
high-mass ellipticals in \hagnn are in
good agreement with the observations while low-mass ellipticals seem to be not
compact enough. 
This effect could be attributed again to limited spatial
resolution, as the size of low-to-intermediate mass galaxies
is only a few times the spatial resolution, and can get some
spurious dynamical support. Therefore, galaxy sizes are
 supposed to converge to $\sim \Delta x=1$ kpc
at the low-mass end, and
it is well plausible that the low-mass galaxies will get more
compact with increased spatial resolution.
Also, it is worth mentioning that the
lensing probability $p$ is a steep function of
the velocity dispersion of galaxies $\sigma$ (i.e. $p \approx \sigma^4$). Then, for a given stellar mass, more compact
galaxies will be selected by using survey which can partly explain the discrepancies between theoretical 
predictions and observations here. In the following, in order to take into account this selection effect, 
 estimations  of the mean value of $\gamma'_{tot}$ are weighted by $\sigma^4$. However, no significant difference 
is noticed in the results and conclusions when $\gamma'_{tot}$ values are derived without this weighting.

%%%%%%%%%%%%%%%%%%%%%%%%%%%%%%%%%%%%%%%%%%%%%%%%%
%     FIG 2 : gamma_tot vs Reff
%%%%%%%%%%%%%%%%%%%%%%%%%%%%%%%%%%%%%%%%%%%%%%%%%
\begin{figure}
\begin{center}
\rotatebox{0}{\includegraphics[width=\columnwidth]{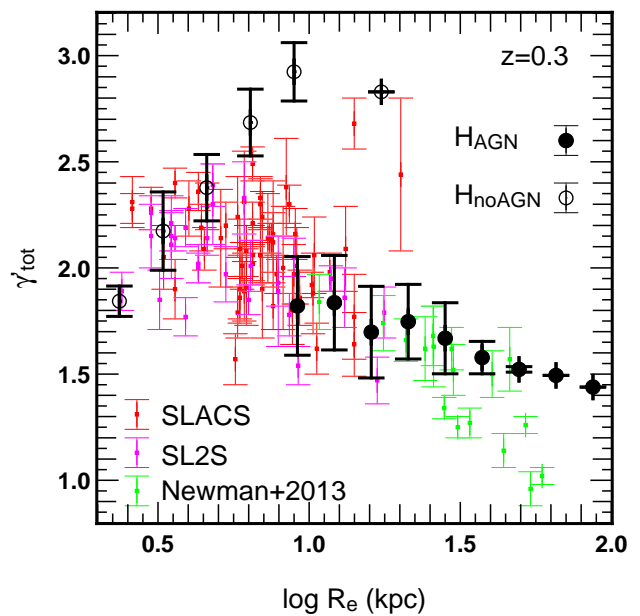}}
\caption{The variations of the mass-weighted total density slopes  $\gamma'_{tot}$ with
respect to the effective radius $R_{e}$. Results are derived from H$_{\rm AGN}$ (black points) and 
 H$_{\rm noAGN}$ (white points) galaxies with $V/\sigma<1$ and a mass of  log$(M_*/M_\odot)>11$ at $z=0.3$.
 We also plot observational data from SLACS (red colors), SL2S (pink colors) and \citet{New++13b} (green colors).
Error bars indicate $1\sigma$ standard deviations. AGN feedback is again required to improve the agreement between theoretical
predictions and observational trends.}
\label{fig2}
\end{center}
 \end{figure}
%%%%%%%%%%%%%%%%%%%%%%%%%%%%%%%%%%%%%%%%%%%%%%%%%

Fig. \ref{fig2} shows now the variations of the mass-weighted total density slope  $\gamma'_{tot}$ with
respect to the effective radius $R_{e}$ at $z=0.3$ for H$_{\rm AGN}$ and H$_{\rm noAGN}$ galaxies
with a mass greater than $10^{11} M_\odot$ and  $V/\sigma<1$.
When comparing our results to observations, we get a much better agreement when AGN feedback 
is taken into account. Indeed, we found that $\gamma'_{tot}$ derived from \hagnn simulation 
is decreasing with $R_{e}$ which is consistent with the observational trend.
On the contrary, an opposite evolution is obtained when AGN  are not included
suggesting that more extended galaxies (or more massive galaxies) tend to have higher $\gamma'_{tot}$ values.
Moreover, for $R_{e} \geq 10$ kpc, the predicted $\gamma'_{tot}$ values derived from the \hagnn simulation 
are in good agreement with those of observations. In particular,  we get  density slopes  close 
to $2$ for the less
massive galaxies of our sample which are values expected from the observations. 
We cannot match the observations for $R_{e}<10$ kpc since H$_{\rm AGN}$ low mass elliptical  galaxies are
too extended as shown in Fig. \ref{fig1}.

Our comparison between the predicted and observed mass-size relation depends also on the assumption
 of the stellar IMF. The data points plotted in Fig. \ref{fig1} are based on a Salpeter IMF. However,
 a recent re-analysis of SLACS lenses shows how the presence of gradients in stellar mass-to-light ratio, 
required by the data, can decrease the inferred stellar masses by as much as $0.2$~dex compared to the
 \citet{Son++15} values \citep{sonnenfeld++18}.
If the values of the stellar masses of SLACS lenses decrease, the data points would move 
in better agreement with the H$_{\rm AGN}$ simulation.

%%%%%%%%%%%%%%%%%%%%%%%%%%%%%%%%%%%%%%%%%%%%%%%%%
%     FIG 3 : gamma_tot vs Mhalos
%%%%%%%%%%%%%%%%%%%%%%%%%%%%%%%%%%%%%%%%%%%%%%%%%
\begin{figure}
\begin{center}
\rotatebox{0}{\includegraphics[width=\columnwidth]{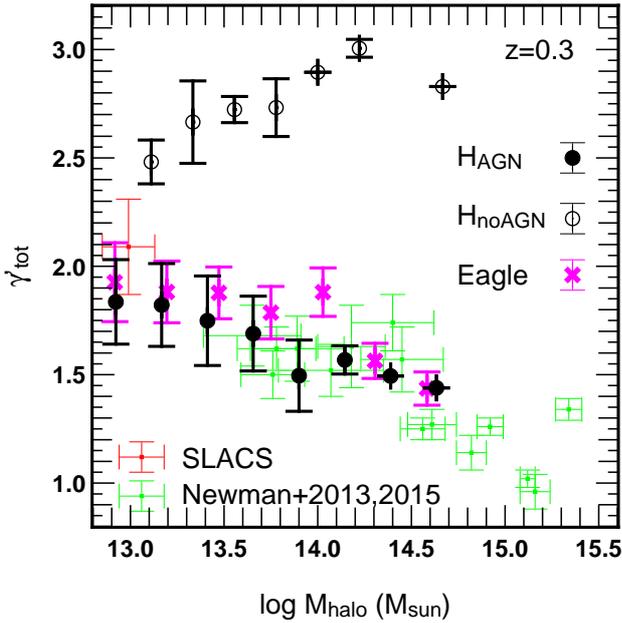}}
\caption{The variations of the mass-weighted total density slopes  $\gamma'_{tot}$ with
respect to the dark matter halo  masses $M_{halo}$. 
Results are derived from H$_{\rm AGN}$ (black points) and  
 H$_{\rm noAGN}$ (white points) galaxies with $V/\sigma<1$ and a mass of  log$(M_*/M_\odot)>11$ at $z=0.3$.
We also plot observational data from SLACS (red colors) and \citet{New++13b,New++15} (green colors)
as well as predictions from the Eagle simulation.
Error bars indicate $1\sigma$ standard deviations. When AGN feedback is included, theoretical predictions are in nice
agreement with observations. On the contrary, in the absence of AGN feedback, derived $\gamma'_{tot}$ values
 are totally inconsistent with observational expectations.}
\label{fig3}
\end{center}
 \end{figure}
%%%%%%%%%%%%%%%%%%%%%%%%%%%%%%%%%%%%%%%%%%%%%%%%%

%%%%%%%%%%%%%%%%%%%%%%%%%%%%%%%%%%%%%%%%%%%%%%%%%
%     FIG 4 : gamma_tot vs M*
%%%%%%%%%%%%%%%%%%%%%%%%%%%%%%%%%%%%%%%%%%%%%%%%%
\begin{figure}
\begin{center}
\rotatebox{0}{\includegraphics[width=\columnwidth]{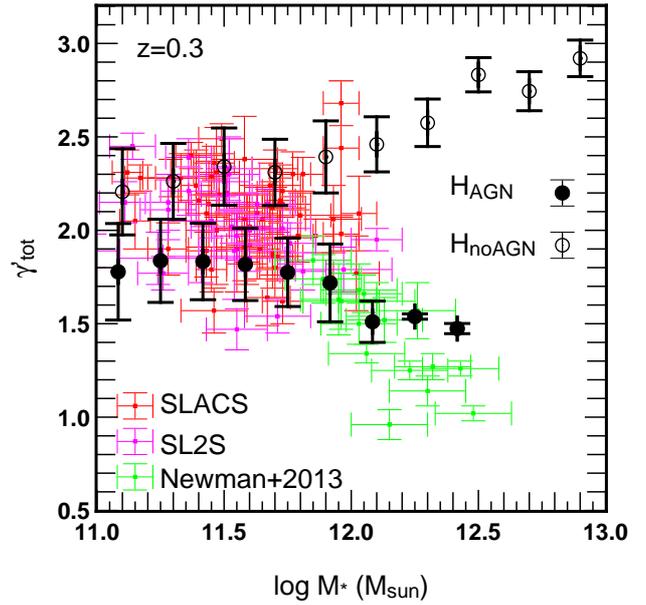}}
\caption{The variations of the mass-weighted total density slopes  $\gamma'_{tot}$ with
respect to the stellar masses $M_{*}$ at $z=0.3$. 
Results are derived from H$_{\rm AGN}$ (black points) and 
 H$_{\rm noAGN}$ (white points) galaxies with $V/\sigma<1$ and a mass of  log$(M_*/M_\odot)>11$.
  We also plot observational data from SLACS (red colors), SL2S (pink colors) and \citet{New++13b} (green colors).
Error bars indicate $1\sigma$ standard deviations. AGN feedback seems to reproduce the observations well.}
\label{fig4}
\end{center}
 \end{figure}
%%%%%%%%%%%%%%%%%%%%%%%%%%%%%%%%%%%%%%%%%%%%%%%%%

From the same samples of galaxies at $z=0.3$ 
(i.e. $M_* \geq10^{11} M_\odot$ and  $V/\sigma<1$), we also take an interest in studying 
the variations of   $\gamma'_{tot}$ with
respect to either the dark matter halo  masses $M_{halo}$ or stellar  masses  $M_{*}$.
Those variations are displayed in Fig.  \ref{fig3}  and   \ref{fig4} respectively.
We find similar trends  to those obtained in Fig. \ref{fig2}. First, 
the presence or not of AGN feedback leads to opposite evolution trends. When AGN feedback is included, 
 $\gamma'_{tot}$ is decreasing with $M_{halo}$ or $M_{*}$. In other words, more massive objects tend
to have more flat total density profiles at the scale of the effective radius. A similar conclusion
was obtained in \cite{Pei++17}, when studying the inner DM and stellar profiles ($r\leq 5$ kpc)
 which is mainly explained
by the fact that AGN feedback has a more important impact in the most massive objects. 
Moreover,  values of  $\gamma'_{tot}$ are  in good agreement
with observational ones. 
On the contrary, in the absence of AGN feedback,  $\gamma'_{tot}$ values are much too high 
 especially for massive objects.
Note that we didn't use our matching scheme here when selecting the H$_{\rm noAGN}$ galaxies
 in order  to consistently compare
with the observational mass range. However, we have checked that the matching scheme 
would select H$_{\rm noAGN}$ galaxies with 
higher masses and higher effective radius  but will  not really change the evolutions
of $\gamma'_{tot}$ previously  derived.  
It is also encouraging to notice that our simulated values for H$_{\rm AGN}$ haloes  presented in Fig. \ref{fig3}
are in good agreement with those of \cite{Schaller++15} using the Eagle simulation \citep{eagle}.
In the Appendix A, we also compare  the results  from lower resolution simulations. We found
good agreement in the different trends relative to the variations of $\gamma'_{tot}$ with
respect to the effective radius, the dark matter halo masses and the stellar masses though 
there are some slight  discrepancies in the variations of the effective radius with respect to 
the stellar masses.

In view of all of these results, AGN feedback seems to be required to explain the observational trends.

\subsection{Dependencies of $\gamma'_{tot}$ with $\gamma'_{dm}$ and $\gamma'_{*}$}

%%%%%%%%%%%%%%%%%%%%%%%%%%%%%%%%%%%%%%%%%%%%%%%%%%%%%%%%%%%%%%%
%     FIG 5 : gamma_tot vs gamma_* and gamma_dm (AGN and noAGN
%%%%%%%%%%%%%%%%%%%%%%%%%%%%%%%%%%%%%%%%%%%%%%%%%%%%%%%%%%%%%%
\begin{figure*}
\begin{center}
\rotatebox{0}{\includegraphics[width=\columnwidth]{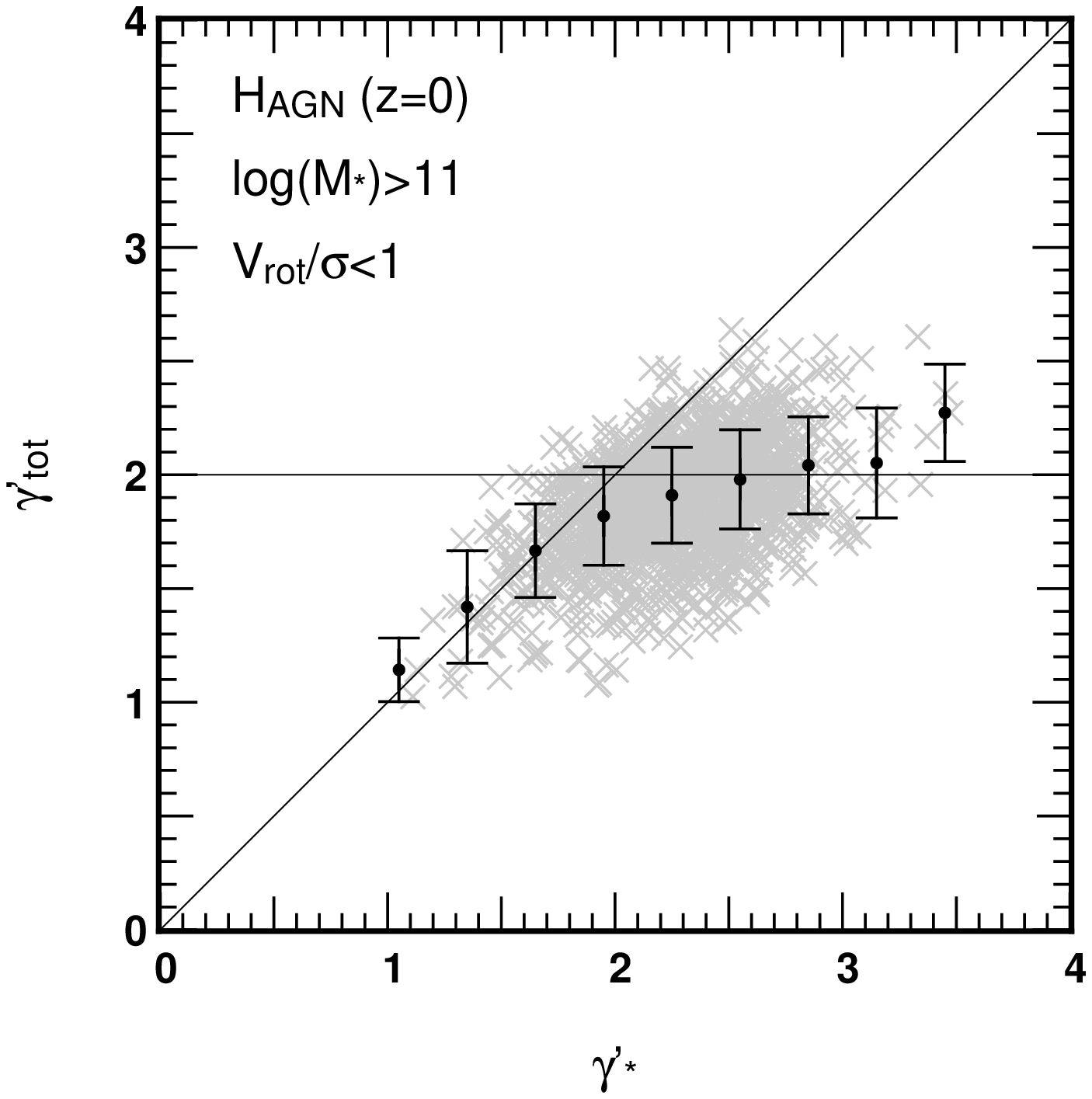}}
\rotatebox{0}{\includegraphics[width=\columnwidth]{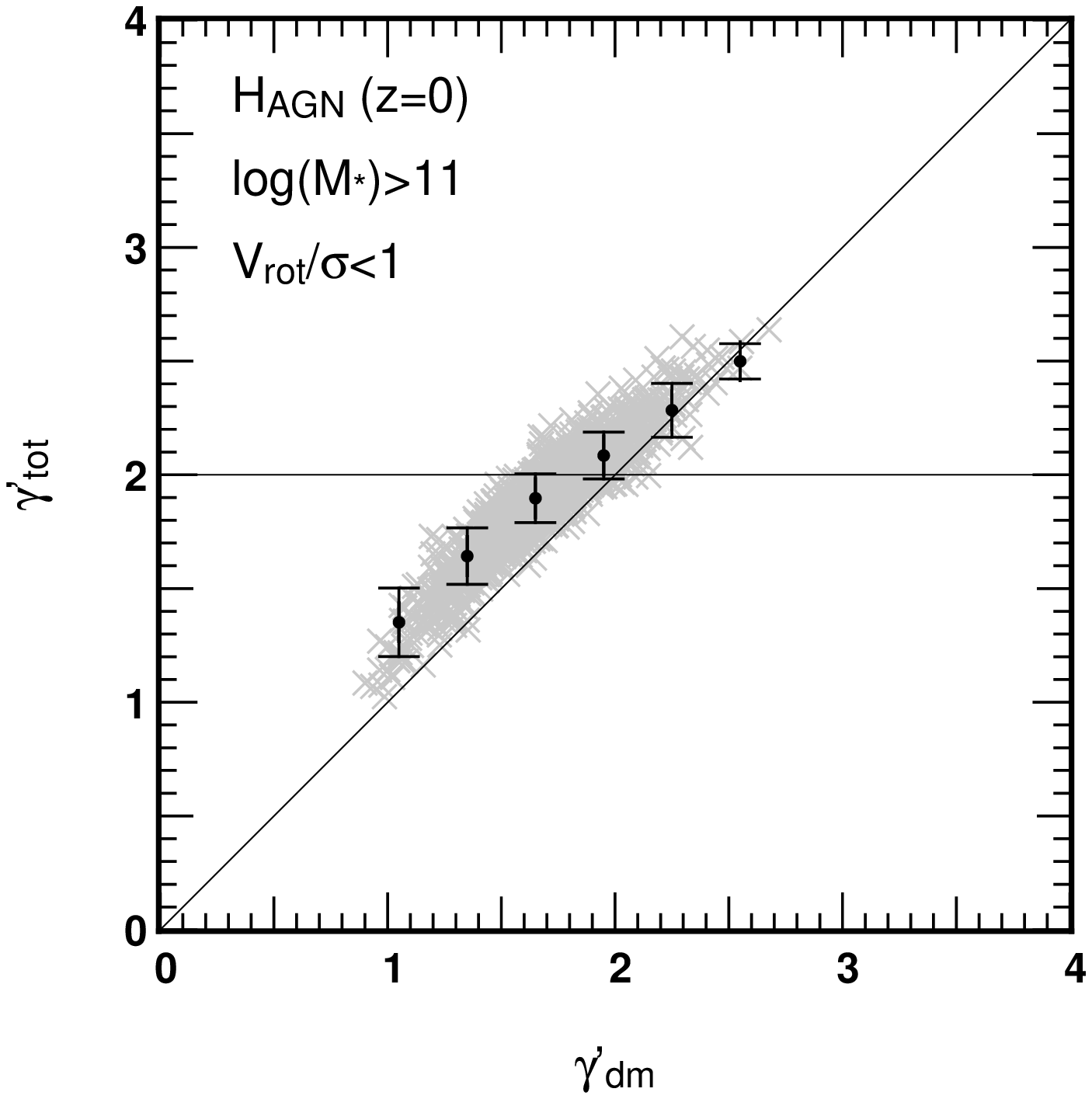}}
\rotatebox{0}{\includegraphics[width=\columnwidth]{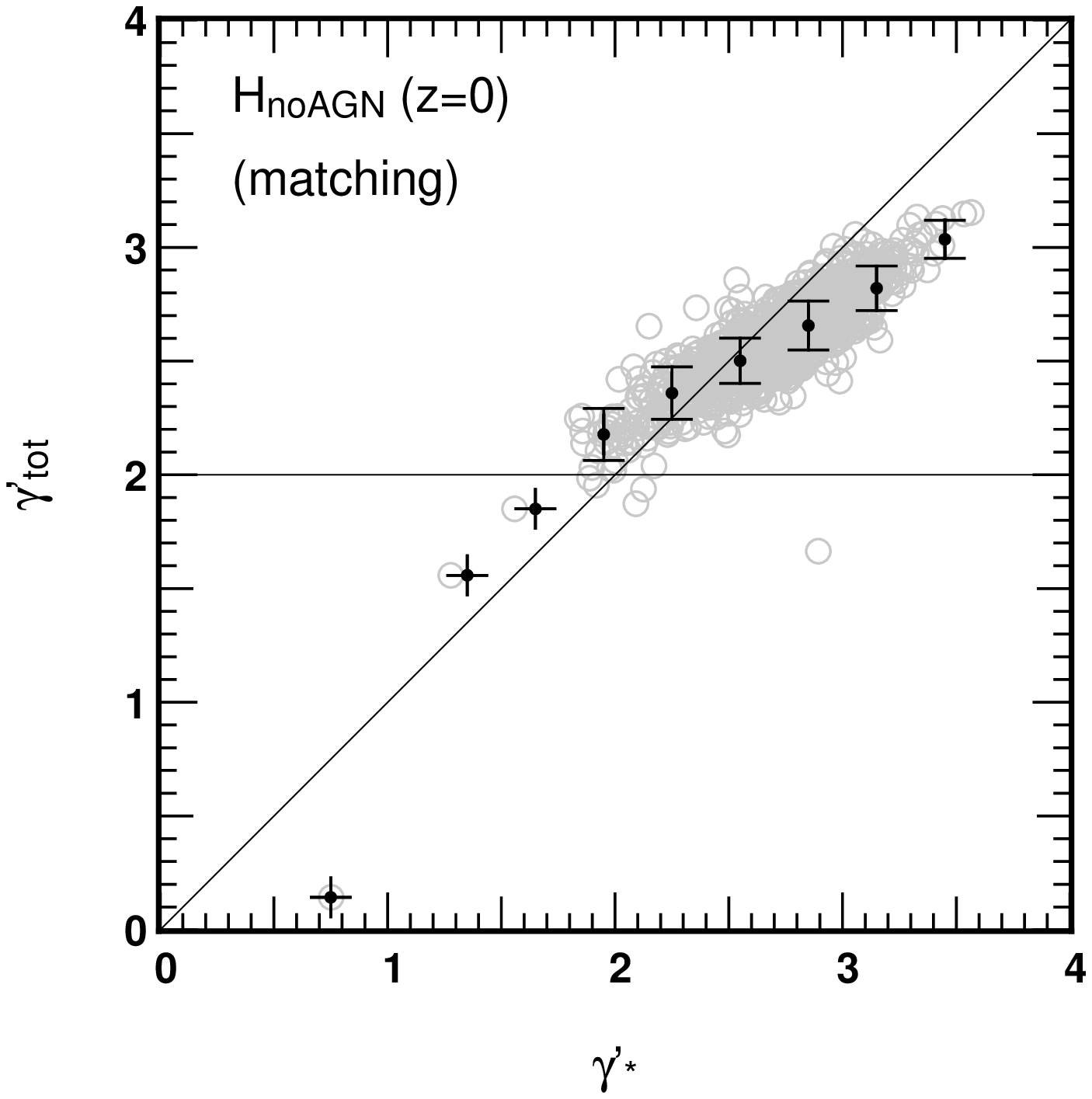}}
\rotatebox{0}{\includegraphics[width=\columnwidth]{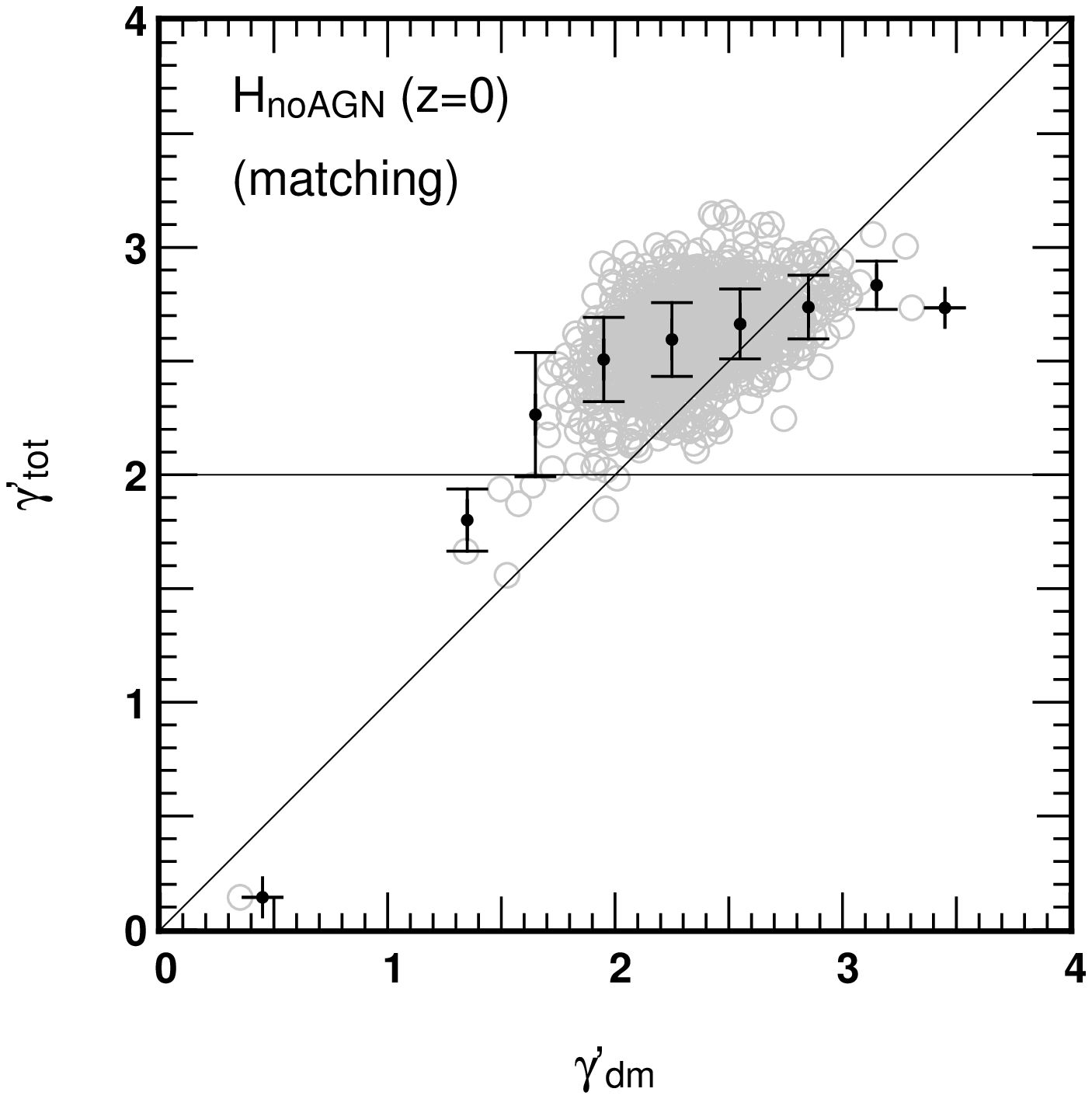}}
\caption{First line: the variations of the mass-weighted total density slope  $\gamma'_{tot}$ with
respect to the mass-weighted stellar density slope  $\gamma'_*$ (first column) or
the mass-weighted dark matter density slope  $\gamma'_{dm}$ (second column) for
 H$_{\rm AGN}$ galaxies with a mass greater than $10^{11} M_\odot$ and $V/\sigma<1$ at
z=0. Results from the  matching H$_{\rm noAGN}$ galaxies are displayed in the second line. The horizontal
solid line represents $\gamma'_{tot}=2$ generally obtained in the observation while the 
diagonal one is $y=x$.}
\label{fig5}
\end{center}
 \end{figure*}
%%%%%%%%%%%%%%%%%%%%%%%%%%%%%%%%%%%%%%%%%%%%%%%%%

In this section, we investigate the variations of $\gamma'_{tot}$
with either the stellar slope $\gamma'_{*}$ or the dark matter slope $\gamma'_{dm}$ 
for galaxies with a mass greater than $10^{11} M_\odot$ and $V/\sigma<1$ at
$z=0.3$. Results from \hagnn and \hnoagnn simulations are presented
in Fig. \ref{fig5}.  First, when AGN feedback is included,
one can notice that $\gamma'_{tot}$ and $\gamma'_{dm}$ are strongly correlated.
Regarding the dependence of  $\gamma'_{tot}$ and $\gamma'_{*}$,  the dispersion is higher and the correlation
is less clear but the important point here is that
AGN feedback tends to limit the total slope to values close to $2$ for
the less massive galaxies while it reduces $\gamma'_{tot}$ in the more massive ones compared to the simulation without AGN,
which is consistent with the observations.
On the contrary, in the absence of AGN feedback, we found again the opposite trends:
$\gamma'_{tot}$ and $\gamma'_{*}$ are this time strongly correlated and 
due to stronger adiabatic contraction, $\gamma'_{dm}$ reach values  too
high in more massive objects \cite[see][]{Pei++17}.

\begin{figure}
\begin{center}
\rotatebox{0}{\includegraphics[width=9cm]{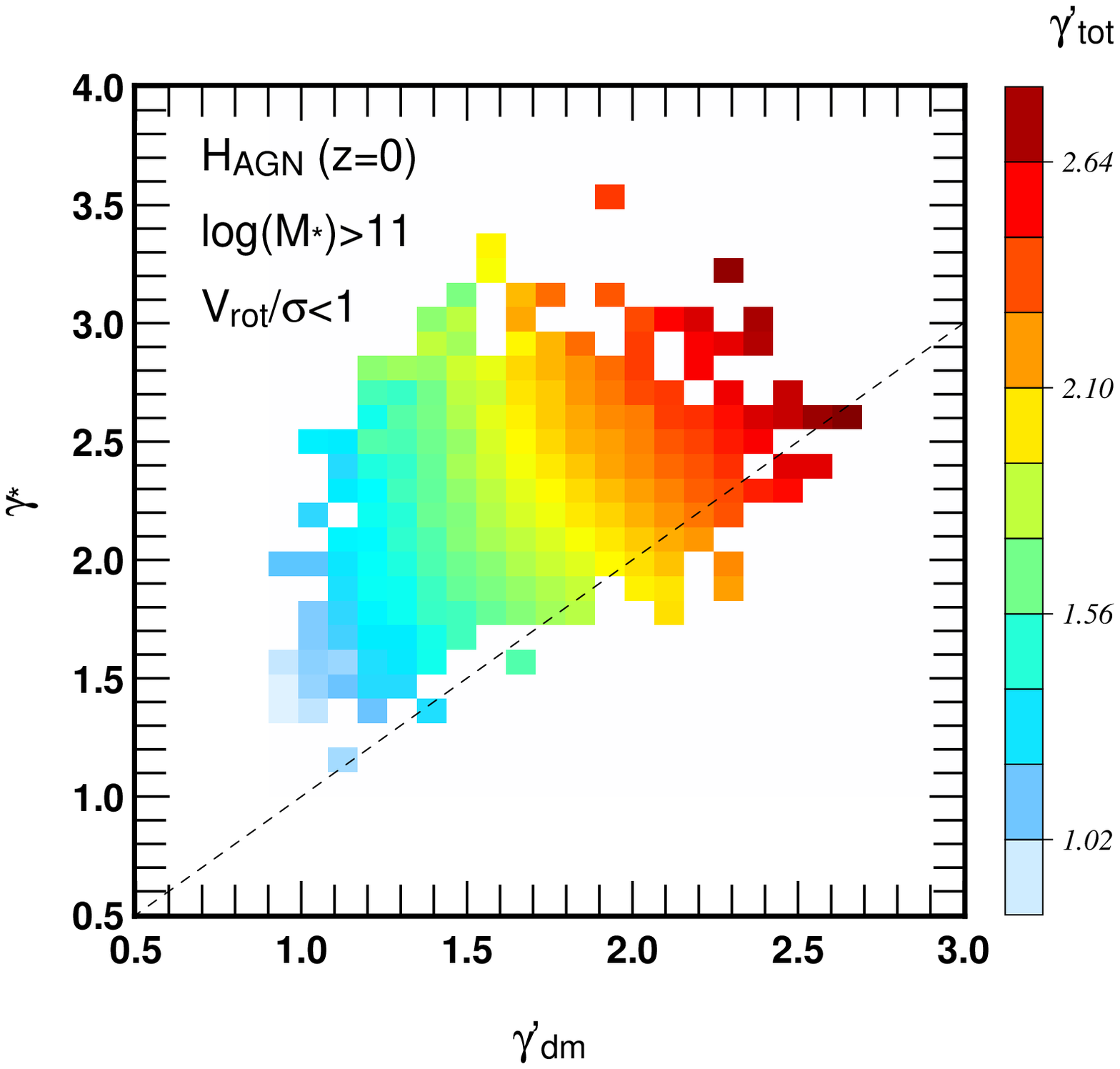}}
\rotatebox{0}{\includegraphics[width=9cm]{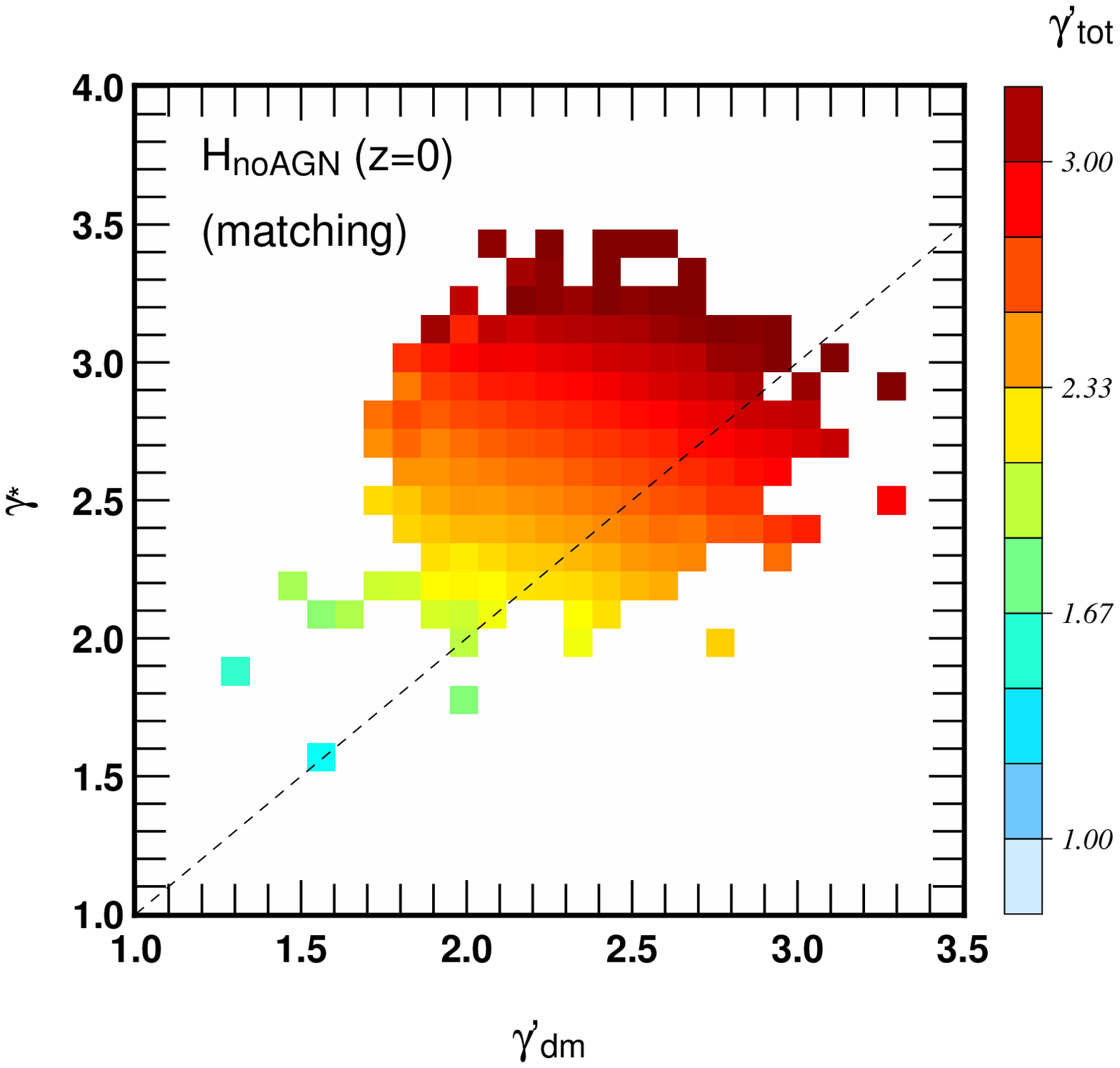}}
%\rotatebox{0}{\includegraphics[width=\columnwidth]{fig17a}}
%\rotatebox{0}{\includegraphics[width=\columnwidth]{fig17b}}
\caption{The variations of the mass-weighted stellar density slope  $\gamma'_{*}$ with
respect to the mass-weighted dark matter density slope  $\gamma'_{dm}$ for
 H$_{\rm AGN}$ galaxies with a mass greater than $10^{11} M_\odot$ and $V/\sigma<1$ at
z=0 (upper panel) and matching  H$_{\rm noAGN}$ galaxies (lower panel). The color
code represents values of the mass-weighted total density slope  $\gamma'_{tot}$.
The dashed line indicates $y=x$. }
\label{fig6}
\end{center}
 \end{figure}

Another way to look into those variations and potential strong correlations between the
three different density slopes
is to consider the two dimensional plots of  Fig. \ref{fig6} showing the variations of $\gamma'_{*}$ and
$\gamma'_{dm}$ with a color code representing the values of $\gamma'_{tot}$ 
for galaxies with a mass greater than $10^{11} M_\odot$ and $V/\sigma<1$ at
$z=0$. Note that this time we use the matching algorithm to select the H$_{\rm noAGN}$ galaxies
%this  to compare the same 
% galaxies bewtween the two simulations. However, we don't see significant differences if
%$h_{noagn}$ galaxies were selected using the same mass and $V/\sigma$ criteria.
which  has the advantage, beside comparing the same objects between the simulations,
to  us to consider a higher number of H$_{\rm noAGN}$ galaxies.
Indeed, most of the massive H$_{\rm noAGN}$ galaxies are disk-dominated and do
 not then satisfied $V/\sigma<1$  \citep[see][]{Dub++16}. However, 
we have checked that no significant  differences   are obtained if
H$_{\rm noAGN}$ galaxies were selected using the same mass and $V/\sigma$ criteria.
Thus, the plots presented in Fig. \ref{fig6} give  the
  possible pairs ($\gamma'_{dm}$,$\gamma'_{*}$) for
a given value of $\gamma'_{tot}$. We also see the linear correlations
between $\gamma'_{tot}$ and $\gamma'_{dm}$
and between $\gamma'_{tot}$ and $\gamma'_{*}$
 when AGN is included or not respectively.

\begin{figure}
\begin{center}
\rotatebox{0}{\includegraphics[width=\columnwidth]{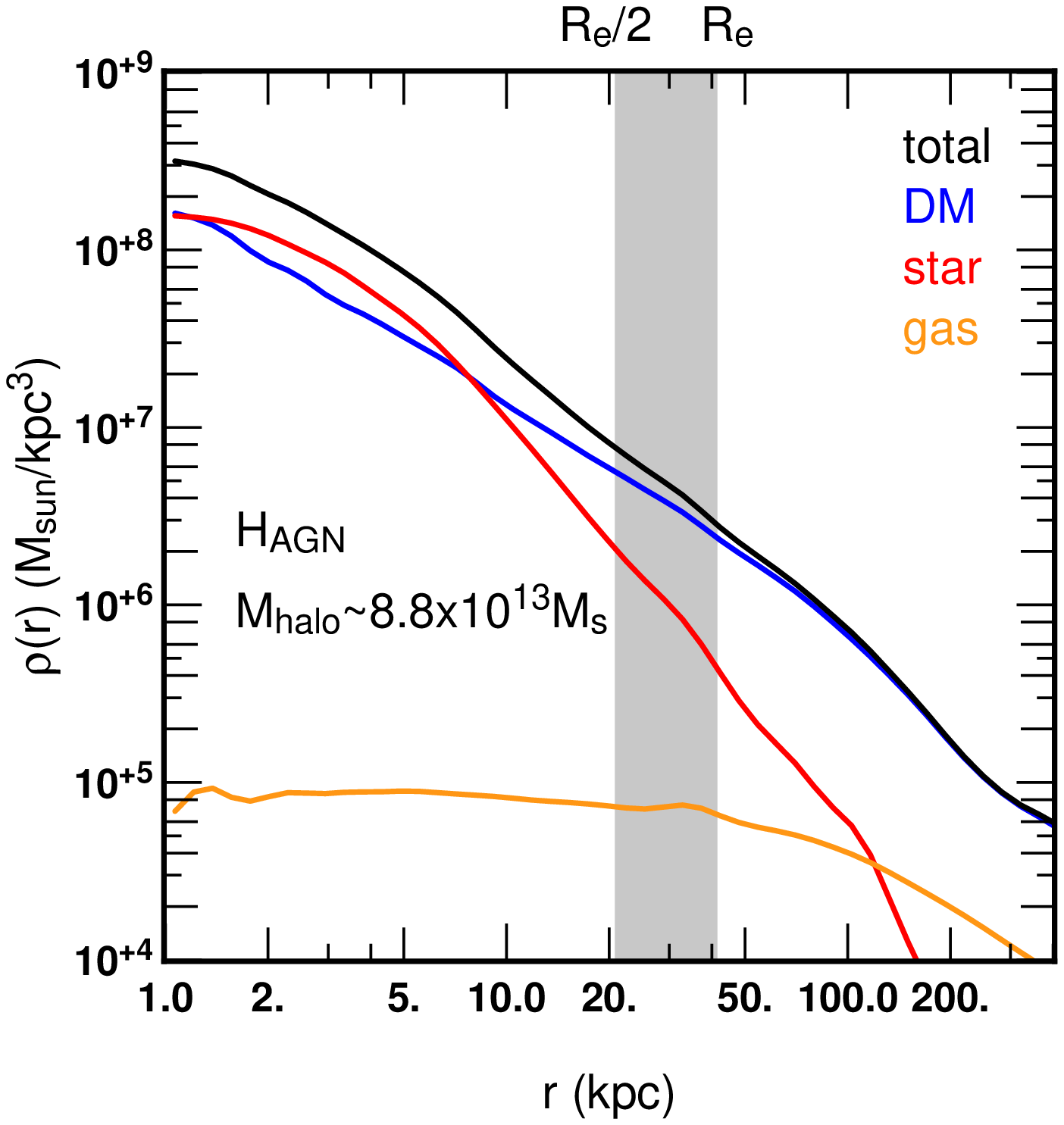}}
\rotatebox{0}{\includegraphics[width=\columnwidth]{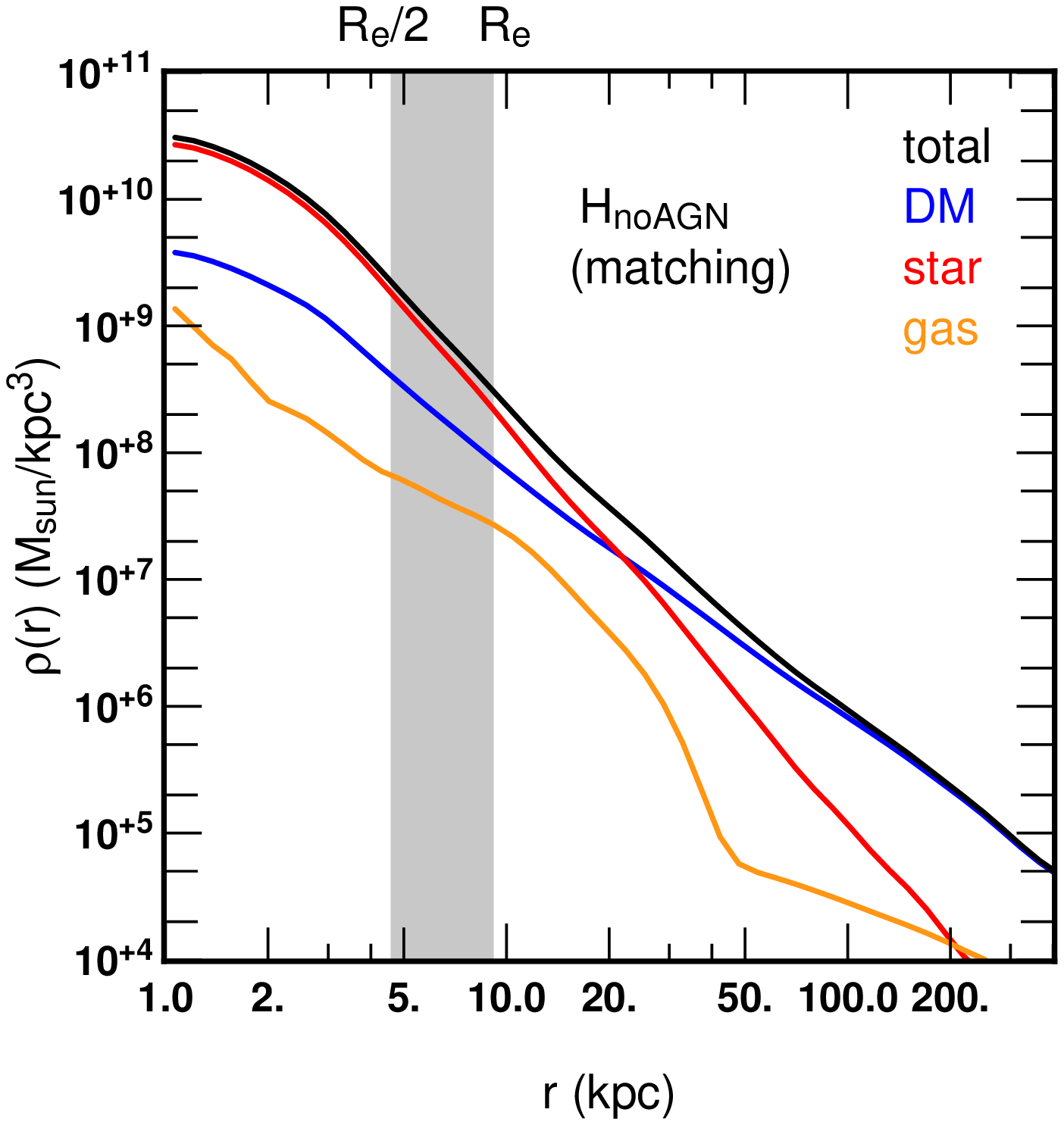}}
%\rotatebox{0}{\includegraphics[width=\columnwidth]{fig17a}}
%\rotatebox{0}{\includegraphics[width=\columnwidth]{fig17b}}
\caption{The total density profile (black lines) of the same massive dark matter halo extracted
from the \hagnn (upper) and \hnoagnn (lower) simulations at $z=0$. The dark matter, stellar and
gas components are also shown in blue, red and orange colors respectively. The grey shade areas
indicate the range of [$R_e/2 - R_e$] i.e. where we compute  $\gamma'_{tot}$, $\gamma'_{*}$ and  $\gamma'_{dm}$.}
\label{fig7}
\end{center}
 \end{figure}

The strong correlations seen between $\gamma'_{tot}$ and $\gamma'_{dm}$ or
$\gamma'_{tot}$ and $\gamma'_{*}$ when AGN is included or not respectively
can be easily understood when studying the density profile of
the different components of a single object shown in Fig. \ref{fig7}.
When AGN  are   included, galaxies are more extended and therefore have  
higher effective  radii in general. Consequently,
the dark matter is the dominant component at the effective radius scale. On the contrary,
galaxies are  found 
%proved 
to be very compact in the absence of AGN activity and have therefore
a smaller effective radius. In this case,  the stellar component is the dominant component
 at the effective radius scale  as
 % it 
 is clearly shown in Fig. \ref{fig7}.

%%%%%%%%%%%%%%%%%%%%%%%%%%%%%%%%%%%%%%%%%%%%%%%%%
%     FIG 6 : gamma_tot vs gamma_* and gamma_dm (2d PLOT) AGN and noAGN 
%%%%%%%%%%%%%%%%%%%%%%%%%%%%%%%%%%%%%%%%%%%%%%%%%
%%%%%%%%%%%%%%%%%%%%%%%%%%%%%%%%%%%%%%%%%%%%%%%%%

%%%%%%%%%%%%%%%%%%%%%%%%%%%%%%%%%%%%%%%%%%%%%%%%%
%     FIG 7 : decompositon of profiles (DM, stars and gas) 
%%%%%%%%%%%%%%%%%%%%%%%%%%%%%%%%%%%%%%%%%%%%%%%%%
%%%%%%%%%%%%%%%%%%%%%%%%%%%%%%%%%%%%%%%%%%%%%%%%%

%%%%%%%%%%%%%%%%%%%%%%%%%%%%%%%%%%%%%%%%%%%%%%%%%
%     FIG 8 : Time evolutions of the gamma_tot vs gamma_* and gamma_dm (STACKED)  
%%%%%%%%%%%%%%%%%%%%%%%%%%%%%%%%%%%%%%%%%%%%%%%%%
\begin{figure}
%\begin{center}
\rotatebox{0}{\includegraphics[width=\columnwidth]{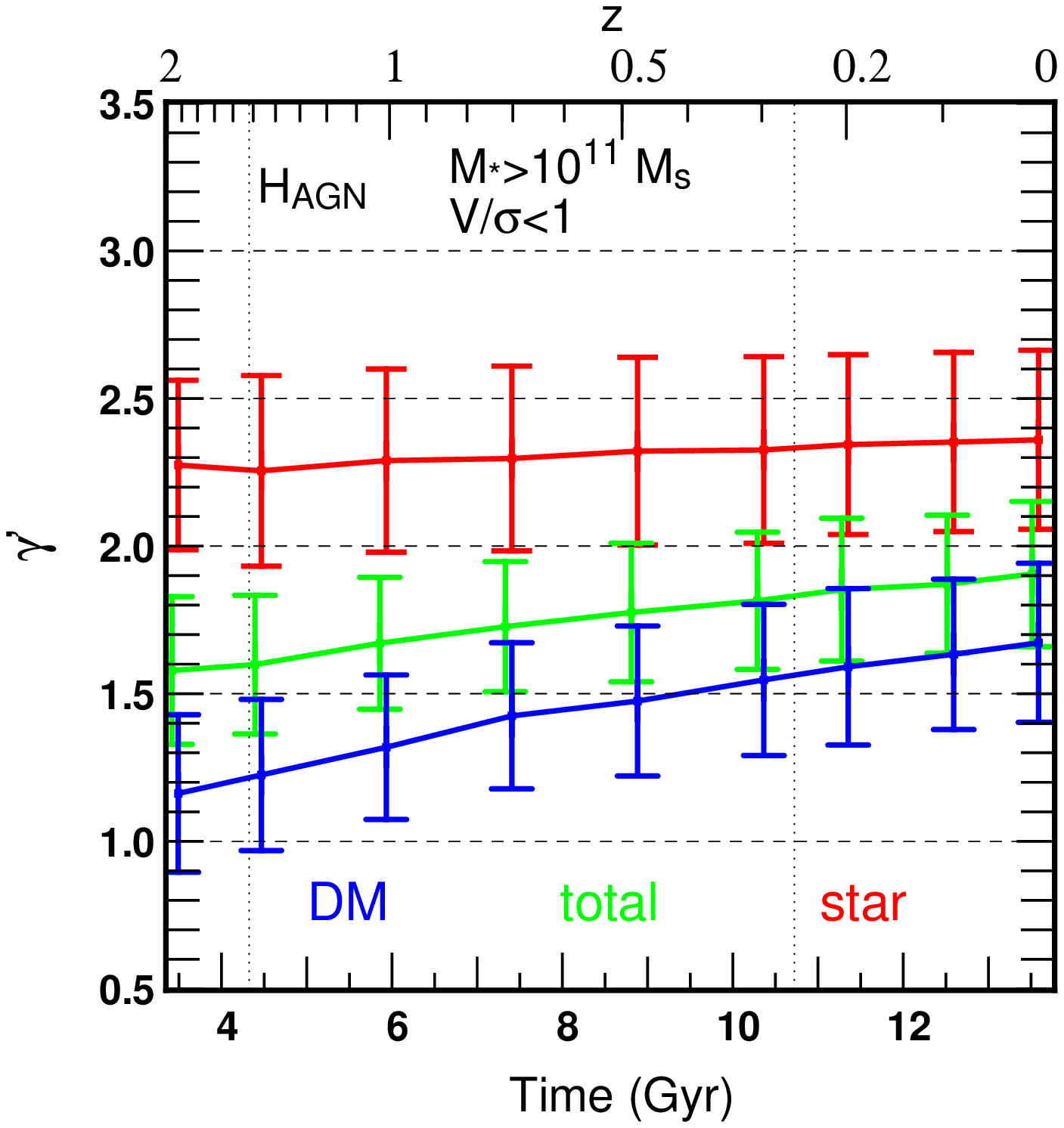}}
%\rotatebox{0}{\includegraphics[width=\columnwidth]{fig_cusp2_8b}}
\rotatebox{0}{\includegraphics[width=\columnwidth]{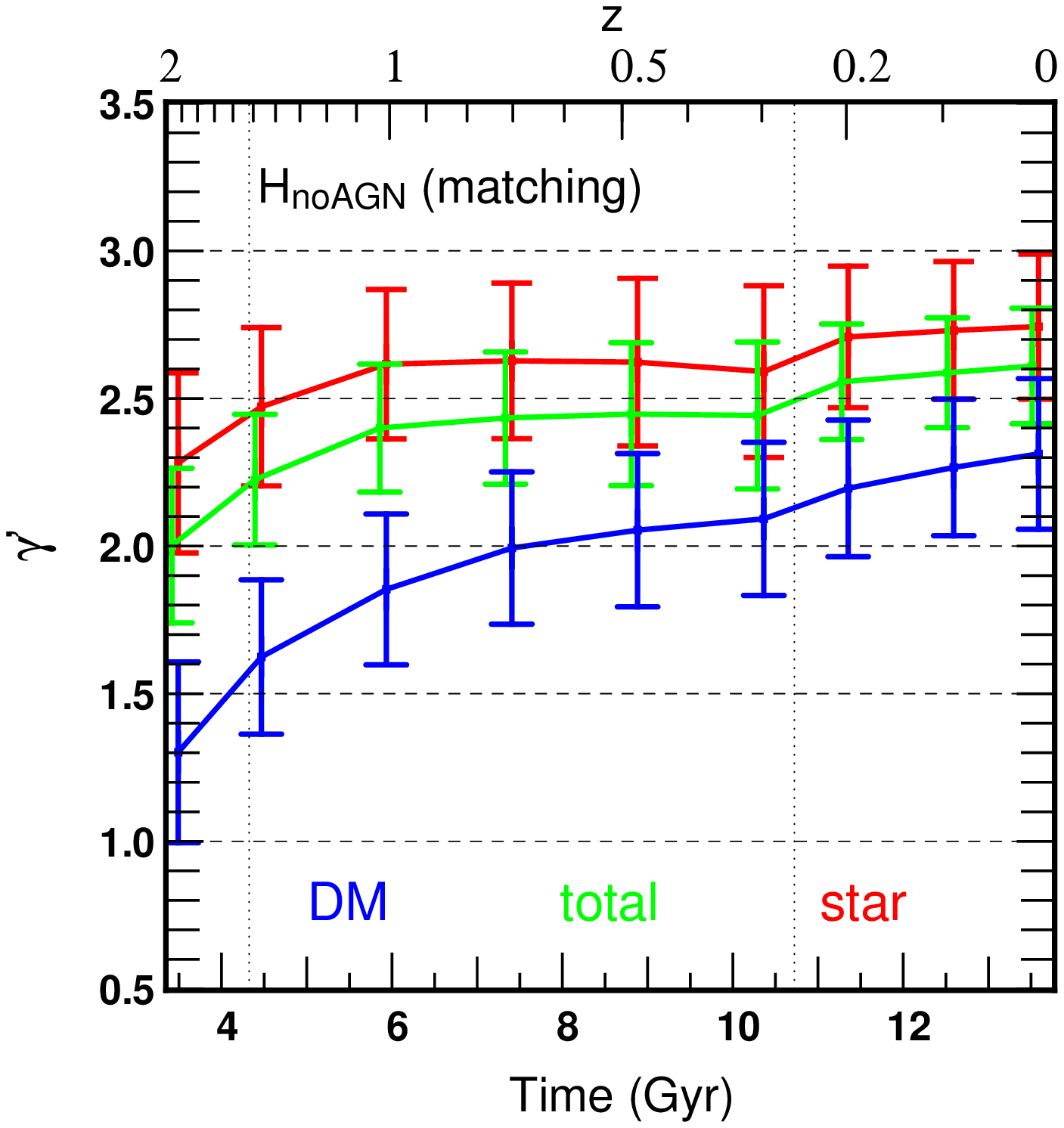}}
%\rotatebox{0}{\includegraphics[width=\columnwidth]{fig_cusp2_8d}}
\caption{First line: the time evolutions of the  mass-weighted total density slope  $\gamma'_{tot}$
 (green lines) for H$_{\rm AGN}$ galaxies with $V/\sigma<1$, a mass greater than $10^{11} M_\odot$ and
$R_e>5$ kpc.
 We also show the variations of the stellar ($\gamma'_{*}$) and
dark matter ($\gamma'_{dm}$) components in red and blue lines respectively. The same evolutions
obtained from the matching  H$_{\rm noAGN}$ galaxies are displayed in the second line. Error bars are
$1\sigma$ standard deviations. The vertical dotted lines indicate the times when a new refinement level is added in the 
simulations. }
\label{fig8}
%\end{center}
 \end{figure}
%%%%%%%%%%%%%%%%%%%%%%%%%%%%%%%%%%%%%%%%%%%%%%%%%

\subsection{ Evolution of $\gamma'_{tot}$ between $0 \leq z \leq 2$}

One of the main objectives of this paper is to make 
theoretical predictions of the
time evolution of $\gamma'_{tot}$ of massive ETGs.
Fig. \ref{fig8} presents the time evolution of $\gamma'_{dm}$, $\gamma'_{*}$ and
$\gamma'_{tot}$ for H$_{\rm AGN}$ galaxies with a mass greater than $10^{11} M_\odot$
and   $V/\sigma<1$ in the redshift range $0 \leq z \leq 2$. We also impose 
$R_e > 5$ kpc in order to be not too close to the lower resolution limit.
We also derived the evolution for   associated  H$_{\rm noAGN}$ galaxies.
Note that we analyse here the
 evolution from $z=2$ since our resolution does not enable us to 
 properly estimate the different density slopes at higher $z$  as 
  typical effective  radii of galaxies at higher redshift become too small. 
 When considering H$_{\rm AGN}$ galaxies,
 we see a slight increase with time of the dark matter and total density slopes while
the stellar density slope is nearly constant and close to $2.3$
in the considered redshift interval. 
 This means that the dark matter and total density profiles estimated
 at the effective radius tend to become slightly steeper at low redshifts
while the stellar one does not vary significantly. 
%It is worth mentioning that at a given redshidt, 
%the mean total slope of the most massive galaxies is in general lower than
%that of the less massive galaxie as already noticed in Fig. \ref{fig4}. 
%This is due to the fact that the total slope of H$_{\rm AGN}$ galaxies
%is strongly correlated to the density slope of the dark matter  
%component which also tend to be lower in the most massive galaxies (see also Fig. \ref{fig3}). 
The situation is slightly different for matching  H$_{\rm noAGN}$ galaxies. In this case,
contrary to $\gamma'_{dm}$ which is increasing,  $\gamma'_{*}$  and therefore $\gamma'_{tot}$
seem to be almost constant after $z=1$.
Note that 
 an additional level of refinement occurs at $z\sim 0.25$. 
 The extra star formation spuriously induced at this epoch
increases the central stellar mass, and induces a
 bump in the evolutions of $\gamma'_{*}$, $\gamma'_{dm}$ (by adiabatic contraction) and therefore $\gamma'_{tot}$. 
 But despite this, it appears clearly that when
 AGN is not taken into account, the density slopes of each component
is always higher than H$_{\rm AGN}$ counterparts.

%The situations is different when studying the evolutions of the density slopes
%of progenitors of $h_{AGN}$  and $h_{noAGN}$ galaxies of mass
%$11<log(M_*)<11.8$ or  $11.9<log(M_*))$ at $z=0$. The different evolutions are shown
%in figure \ref{fig8}. When AGN is included, the dark matter, stellar and total density slopes
%are continuously increasing for the mass sample of $11<log(M_*)<11.8$. For
%the most massive galaxies ($11.9<log(M_*))$ at $z=0$), the different density slopes
%are increasing until $z\sim 0.7$ and remains constant afterwards.
% As far as the $h_{noAGN}$ galaxies are concerned, the different density profiles, 
% after increasing from $z=2$ to $z\sim 0.7$ tend to remains constant afterward
% in spite of the sudden bump in their evolutions induced by a new 
% level of refinement.
 
\begin{figure}
\begin{center}
\rotatebox{0}{\includegraphics[width=\columnwidth]{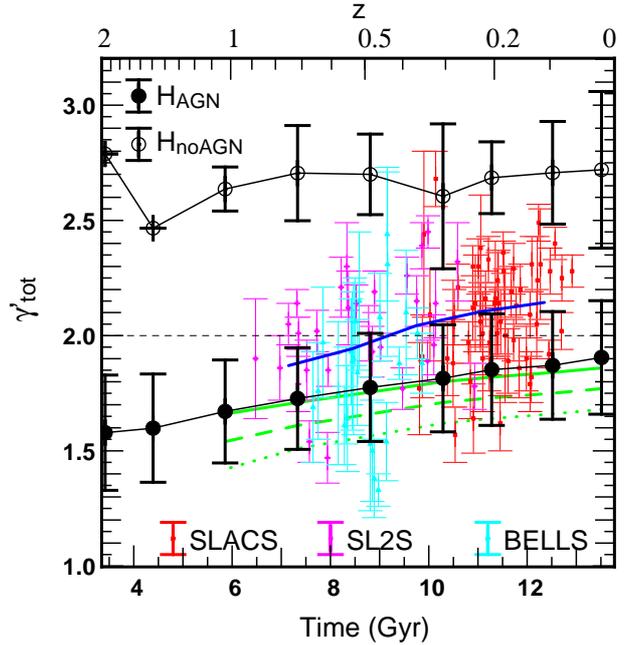}}
\caption{The time evolutions of the  mass-weighted total density slope  $\gamma'_{tot}$
 for H$_{\rm AGN}$ (black points) and H$_{\rm noAGN}$  (white points) galaxies 
satisfying  $V/\sigma<1$, $M_* \geq 10^{11} M_\odot$ and $R_e>5$ kpc in the considered redshift interval.
 We also add the observational data
from the SLACS (red), SL2S (purple) and BELLS (cyan). The solid blue line shows the mean evolution
derived from all observational data. 
The solid, dashed and dotted green lines are derived from respectively the 84, 50 and 16 percentiles of the probability 
distribution function of $<\gamma'_{tot}>$ from Equation (2). They represent the mean evolutions of $\gamma'_{tot}$ 
after removing the dependence on stellar mass density (see details in the text). Error bars are $1\sigma$ standard deviations.}
\label{fig9}
\end{center}
 \end{figure}

 Finally, we directly compare the  evolution of $\gamma'_{tot}$ to observations in
 Fig. \ref{fig9}. We consider here again galaxies satisfying 
$M_* \geq 10^{11} M_\odot$,  $V/\sigma<1$  and $R_e > 5$ kpc in the redshift range of $0 \leq z \leq 2$. In the absence of AGN feedback,
the derived $\gamma'_{tot}$ values are too high compared to observations, as expected.
When AGN feedback is included, the derived $\gamma'_{tot}$ values
are this time  slightly
%a bit 
too low.
The observational data suggest that $\gamma'_{tot}$ is also slightly increasing in the interval $0\leq z \leq 0.8$
though there is a  large
%big 
dispersion in the data.
 Our theoretical 
prediction from \hagnn exhibits 
%then  a 
much better agreement but it is not yet
fully satisfactory. The discrepancies with observations could be explained by
the fact that the H$_{\rm AGN}$ galaxies are
too extended as shown in Fig. \ref{fig1}. Having galaxies slightly more compact,
especially for the low mass ellipticals, will tend to increase values of $\gamma'_{tot}$ 
and therefore improve greatly 
%the 
matching with observations.

For a more accurate comparison between our predictions and observations, we need to take into 
account the additional dependence of $\gamma'_{tot}$ on stellar mass and size.
In other words, at each $z$, we need to compare the average $\gamma'_{tot}$ of our simulated
 galaxies with the average $\gamma'_{tot}$ observed for galaxies with the same values of $M_*$ and $R_e$.
The latter can be obtained by evaluating Equation (2) at the same $M_*$ and $R_e$ as the simulation average.
This gives us a band  which is in general different from the simple average over data points.
In Fig \ref{fig9}, we plot the curves relative to the 16, 50 and 84 percentile of the distribution (green lines).
One can notice that, after removing the dependence on mass and size, $<\gamma'_{tot}>$ is still increasing
and therefore this increase should not be due to the size-redshift evolution of ETGs galaxies.

The slight increase of $\gamma'_{tot}$ in the considered redshift interval seems to be due to
the increase of the density slope of the DM component as suggested by Figs \ref{fig5}, \ref{fig6} and \ref{fig8}.
 As shown in  \cite{Pei++17}, 
the evolution of the inner part of \hagnn haloes exhibit
 a condensation phase or ``cusp regeneration''  from $z\sim 1.6$ down to $z=0$. 
This phase is associated with a decrease of the evolution of the  mass accretion on to black holes (BHs)
 $\dot{M}_{\rm BH} \equiv dM_{\rm BH}/dt$ and therefore the AGN activity (see Fig 5 in \cite{Pei++17}).
Indeed, as advocated by \cite{peirani+08} and \cite{martizzietal12clumps},
repetitive cycles of gas expansion by
AGN feedback (preferably through quasar mode) and gas cooling can efficiently flatten the inner part of dark matter profiles.
If the AGN activity progressively decreases, 
this proposed mechanism becomes less efficient at counterbalancing the DM adiabatic contraction 
and  at keeping the DM density profiles flat.
Note that the density slope of the DM component in the \hnoagnn run is also increasing 
due to the same physical effect namely the adiabatic contraction. However, this not 
really affect the evolution of the total density profile as this latter
mainly depends on the evolution of the stellar component.

\begin{figure}
\begin{center}
\rotatebox{0}{\includegraphics[width=\columnwidth]{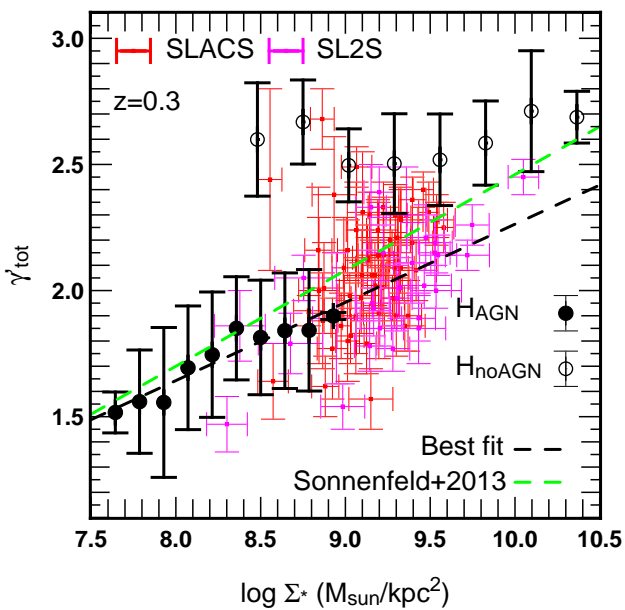}}
\caption{The variations of the mass-weighted total density slopes  $\gamma'_{tot}$ 
as a function of stellar mass density at $z=0.3$. We consider here the same samples of simulated galaxies than
previously. The black dashed line is the best fit measuring the inference on the parameter describing 
the dependence on $\Sigma_*$ namely $\eta = 0.31 \pm 0.04$. The green dashed one indicates 
$\eta = 0.38 \pm 0.07$ derived from observations \citep[see][]{PaperIV}.
}
\label{fig11}
\end{center}
 \end{figure}

\cite{PaperIV} have shown that the dependence of $\gamma'_{tot}$
 on the structure of ETGs can be well summarized with a dependence on stellar mass density
 $\Sigma_* = M_*/(2\pi R^2_{e})$, leaving little dependence on $M_*$ and $R_e$ individually. 
Equation (2) can then be simplified as follows:
 \begin{equation}
 <\gamma'_{tot}> = \gamma'_{0} + \alpha(z-0.3) + \eta(log \Sigma_* - 9).
 \end{equation}
\citet{PaperIV} measured a value of $\eta=0.38\pm0.07$ from the sample of SLACS and SL2S lenses.
Our H$_{\rm AGN}$ model is able to reproduce the observed trend with $\Sigma_*$, as shown in Fig.~\ref{fig11}.
 By fitting our distribution of $\gamma'_{tot}$ as a function of $\Sigma_*$ for galaxies at $z=0.3$, 
we find $\eta = 0.31 \pm 0.04$, in good agreement with the \citet{PaperIV} value.
But once again, since the  H$_{\rm AGN}$ galaxies tend to be too extended, the
 derived $\Sigma_*$ values are too small compared to the observational values from the same galaxy mass range.
This raises an additional question: we have used Equation (2) to estimate the observed $\gamma'_{tot}$ for
 galaxies with the same mass and size as the simulated ones, however there is little overlap in stellar
 mass density between the SLACS and SL2S lenses and the H$_{\rm AGN}$ galaxies. The use of Equation (2)
 for our simulation is then, strictly speaking, an extrapolation. Nevertheless, \citet{New++15}
 showed how a very similar correlation between $\gamma'_{tot}$ and $\Sigma_*$ holds down to $\log{\Sigma_*}=8.0$.
Our use of Equation (2) in the stellar mass density regime probed by the H$_{\rm AGN}$ galaxies is
 then supported by observations.

%%%%%%%%%%%%%%%%%%%%%%%%%%%%%%%%%%%%%%%%%%%%%%%%%

%%%%%%%%%%%%%%%%%%%%%%%%%%%%%%%%%%%%%%%%%%%%%%%%%
%     FIG 11 : gamma_tot vs S*
%%%%%%%%%%%%%%%%%%%%%%%%%%%%%%%%%%%%%%%%%%%%%%%%%
%%%%%%%%%%%%%%%%%%%%%%%%%%%%%%%%%%%%%%%%%%%%%%%%%

\section{Discussion and Conclusions}\label{sec:disc}

By comparing results from two state-of-the-art hydrodynamical
 cosmological simulations whose only difference is the
 presence/absence of AGN  feedback,
we have explored the impact of AGN feedback on the
evolution of the total density profiles of massive early-type galaxies. 
We mainly focused on galaxies with a mass greater than  $10^{11} {\rm M}_\odot$
and satisfying $V/\sigma<1$. Our findings can be summarized as follows:

$\bullet$  
In the absence of AGN feedback,
the simulated galaxies are clearly too compact with an effective radius too small,
 compared to observational data.
On the contrary, AGN feedback tends to form more extended galaxies 
as already noted by \cite{Dub++13,Dub++16}. In this case,
high-mass ellipticals in \hagnn seem to be in
good agreement with the observations while low-mass ellipticals are not
compact enough.

$\bullet$  
When studying the variations of $\gamma'_{tot}$  with the effective radius, 
the galaxy mass and the host halo mass  at  $z=0.3$, we found that the inclusion of AGN feedback
is required to get satisfactory agreement with observational values and trends.

$\bullet$
$\gamma'_{tot}$ is strongly correlated with $\gamma'_{dm}$ when AGN feedback is included. On
the contrary, $\gamma'_{tot}$ is strongly correlated with $\gamma'_{*}$ in
the absence of AGN feedback. 

$\bullet$
$\gamma'_{tot}$ is slightly increasing between  $0\leq z \leq 2$. This is due to the fact
that when AGN is included, the evolution of $\gamma'_{tot}$ is correlated with  
the evolution of $\gamma'_{dm}$. In \cite{Pei++17}, we indeed found that
the dark matter density slope is increasing at low redshift because
the AGN activity is reduced. 

%An important  issue that our simulations do not address  is the inefficiency of AGN feedback. High resolution simulations
% of individual systems  \cite{costa14} show that outflows are highly anisotropic and hence inefficient.
% This can only increase for lower mass systems. A plausible possibility is that lower mass systems are 
%less affected by AGN feedback, hence retaining steeper profiles.

One interesting prediction of the present analysis is the evolution
of $\gamma'_{tot}$ over 
%the 
cosmic time. Because of our limited resolution (1 physical kpc),
we could only consider the interval $0\leq z \leq 2$. When AGN
 are  included,
we found that $\gamma'_{tot}$ is slightly increasing with time. Compared to observational values,
this trend is quite consistent,  although the simulated values are slightly too low. 
This could be explained by the fact that  
the less massive H$_{\rm AGN}$ galaxies in our samples seem to be too extended.
It is worth mentioning that observations from strong lensing by
 \cite{ruff2011} and \cite{Bol++12}
 suggest that the total density profile of massive galaxies has become slightly steeper over cosmic time,
in agreement with our findings.
However, using well resolved hydrodynamical simulations, \cite{remus17,Xu++16} found the opposite trend i.e, 
 $\gamma'_{tot}$ is 
decreasing after $z=2$.
Probably, the key point here is  the behaviour  and values of $\gamma'_{tot}$  before $z=2$. If important sources of 
feedback (e.g. AGN quasar mode) can flatten the dark matter and stellar components
at high redshift \citep[e.g.][]{peirani+08,martizzietal12clumps}, then it would be difficult to sustain such flat profiles at lower redshift
if feedback become less efficient as advocated by \cite{Pei++17}. On the contrary,
if these different components are already steep at  $z=2$, then mechanisms such as (dry) major mergers \citep{SNT14} or
efficient feedback could more easily  flatten them at low redshift.
The observational data of the SLACS, SL2S and BELLS seems to suggest that $\gamma'_{tot}$
is slightly increasing but future detailed observational data, especially before $z=2$, will definitely help to 
 constrain the different scenarios and theoretical models.

\vspace{1.0cm}

\noindent
{\bf Acknowledgements}

\noindent
We are grateful to the referee Alan R.\, Duffy {\bf and the editor Joop Schaye} for
giving constructive comments which substantially helped improving the quality of the paper.
We warmly thank M.\,Schaller for providing relevant Eagle simulation data.
S.\,P. acknowledges support from the Japan Society for the
 Promotion of Science (JSPS long-term invitation fellowship).
This  work  was  granted  access  to  the  HPC  resources  of
CINES under the allocations 2013047012, 2014047012 and
2015047012  made  by  GENCI and  has  made  use
of  the  Horizon  cluster hosted by the Institut d'Astrophysique de Paris
on  which  the  simulation  was post-processed.
This work is supported in part by JSPS KAKENHI Grant Number JP26800093, JP15H05892 and JP17K14250
and by World Premier International Research Center Initiative (WPI Initiative), MEXT, Japan.
The research of J.\,D. is supported by Adrian Beecroft and STFC.
The research of J.\,S. has been supported at IAP by ERC project 267117 (DARK) hosted by Universit\'e Pierre et Marie Curie - Paris 6.
This work was carried out within the framework of the
Horizon project (\texttt{http://www.projet-horizon.fr})
and is partially supported by the grants ANR-13-BS05-0005 
of the French Agence Nationale de la Recherche.

\bibliographystyle{aa}
\bibliography{ref}

%%%%%%%%%%%%%%%%% APPENDICES %%%%%%%%%%%%%%%%%%%%%

\appendix

\section{Resolution tests}

In order to investigate the effects of resolution on our results,
we compare our fiducial simulations (i.e. \hagn) with lower resolution
versions (i.e. 512$^3$ and 256$^3$ dark matter particles with initial uniform grid
refined down to $\Delta x=2$ and $\Delta x=4$ proper kpc  respectively).
To have a clear diagnostic,
% we follow the same strategy than for instance in\cite{duffy++10}, namely 
we focus on the average density profiles of dark matter haloes 
at two epochs, $z=1$ and $z=0.3$.
In order to match dark matter haloes and galaxies between the three simulations, we use the same 
scheme developed in \cite{Pei++17} and summarized in section 2.2. 
The main difference here is that one given dark matter particle of any of the two lower resolution simulations
is associated to 8 particles  in the subsequent higher resolution
version (i.e. between \hagnn and \hagn-512$^3$
or between \hagn-512$^3$ and \hagn-256$^3$). However, it is still possible to know the composition
of each dark matter halo and the fraction of common particles between two objects of different simulation versions.
If this fraction is greater than 50\%, we consider that these DM haloes are twin.

In the following, we select galaxy samples of a  given mass range in the \hagnn simulation
and at $z=1$ or $z=0.3$. We then identified their matching galaxies in 
  \hagn-512$^3$ and \hagn-256$^3$ when possible. Finally, we identify their respective host halo 
to produce samples of DM haloes between the three simulations. Only  dark matter haloes which can be matched between the three
simulations are kept. %This means that for a given stellar mass range, each sample of host dark matter haloes
% used to compute the mean density profile in the following have the same number of objects.

\begin{figure}
\begin{center}
\rotatebox{0}{\includegraphics[width=\columnwidth]{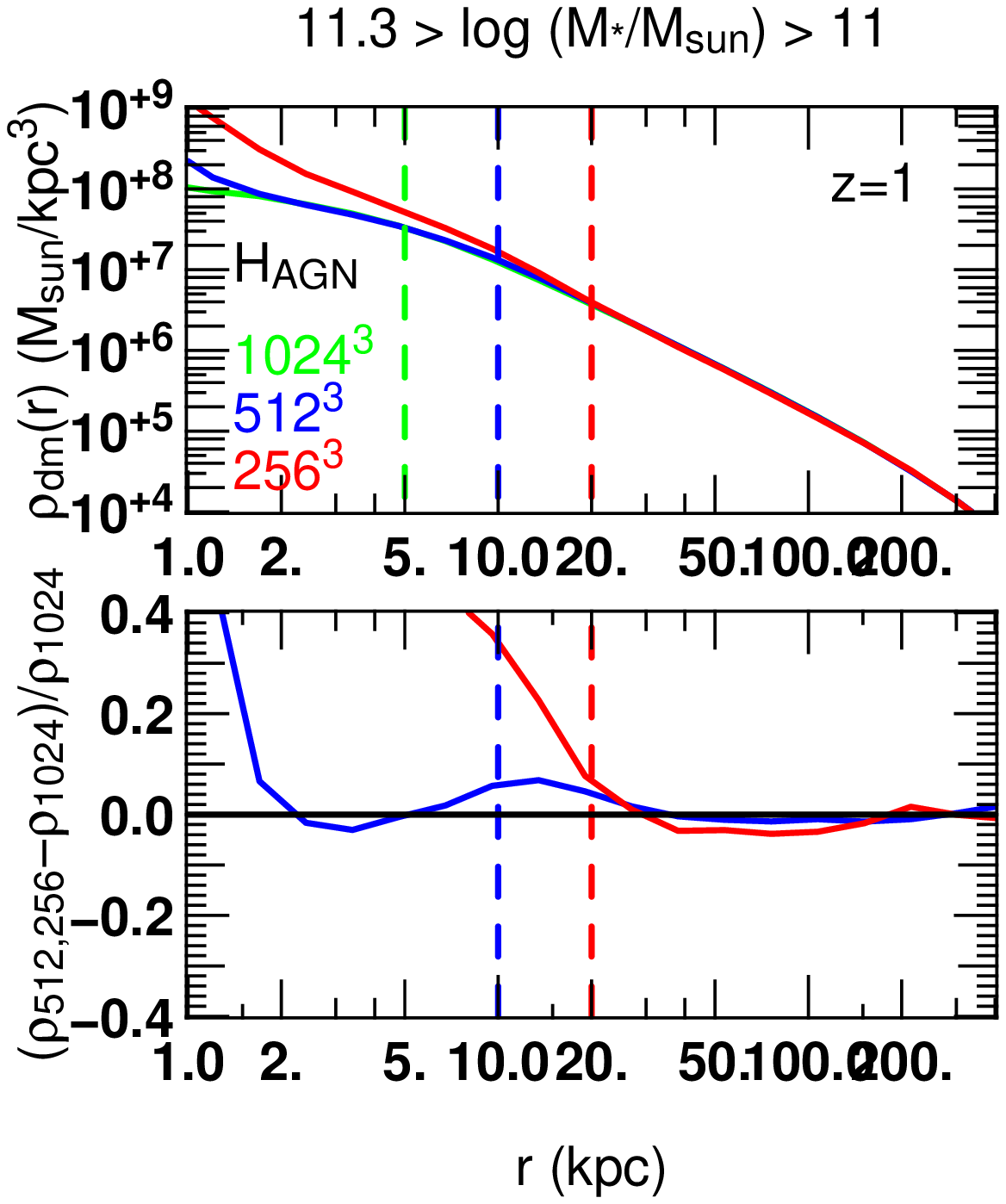}}
\caption{Comparison of the mean dark matter density profiles (and residual) of haloes hosting a galaxy
with a mass of $ 2\times 10^{11}\geq M_* \geq 10^{11} M_\odot$ at $z=1$, derived from
the \hagnn simulation (green line) and lower resolution versions, \hagn-512$^3$ (blue line) and
\hagn-256$^3$ (red line). The dashed vertical lines (with same color code)  correspond to recommended lower limit values suggested
by Power et al. (2003) for each simulation which  give satisfactory indication of where 
the density profiles should converge.}
\label{figapp1}
\end{center}
 \end{figure}

\begin{figure*}
\begin{center}
\rotatebox{0}{\includegraphics[width=\columnwidth]{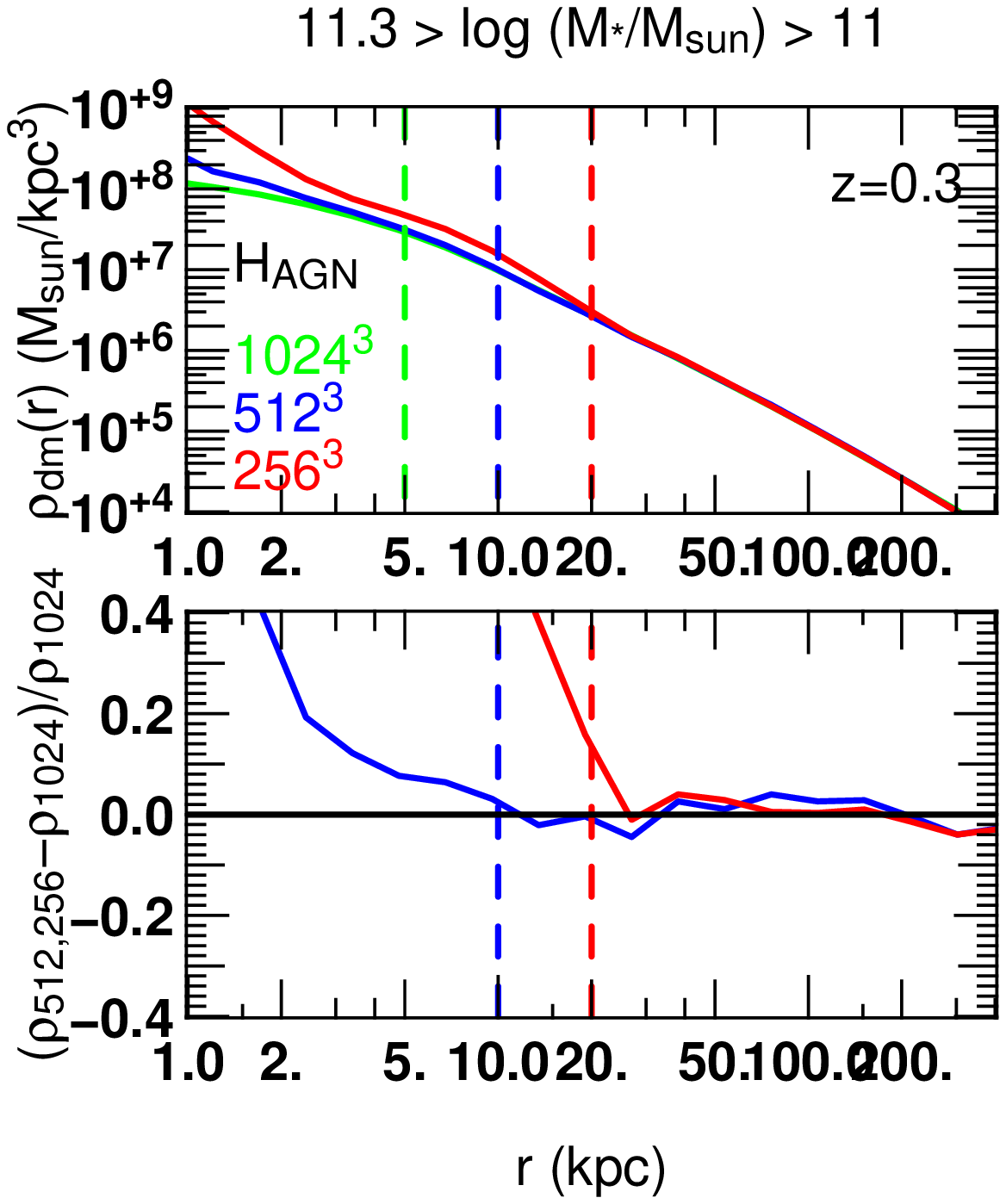}}
\rotatebox{0}{\includegraphics[width=\columnwidth]{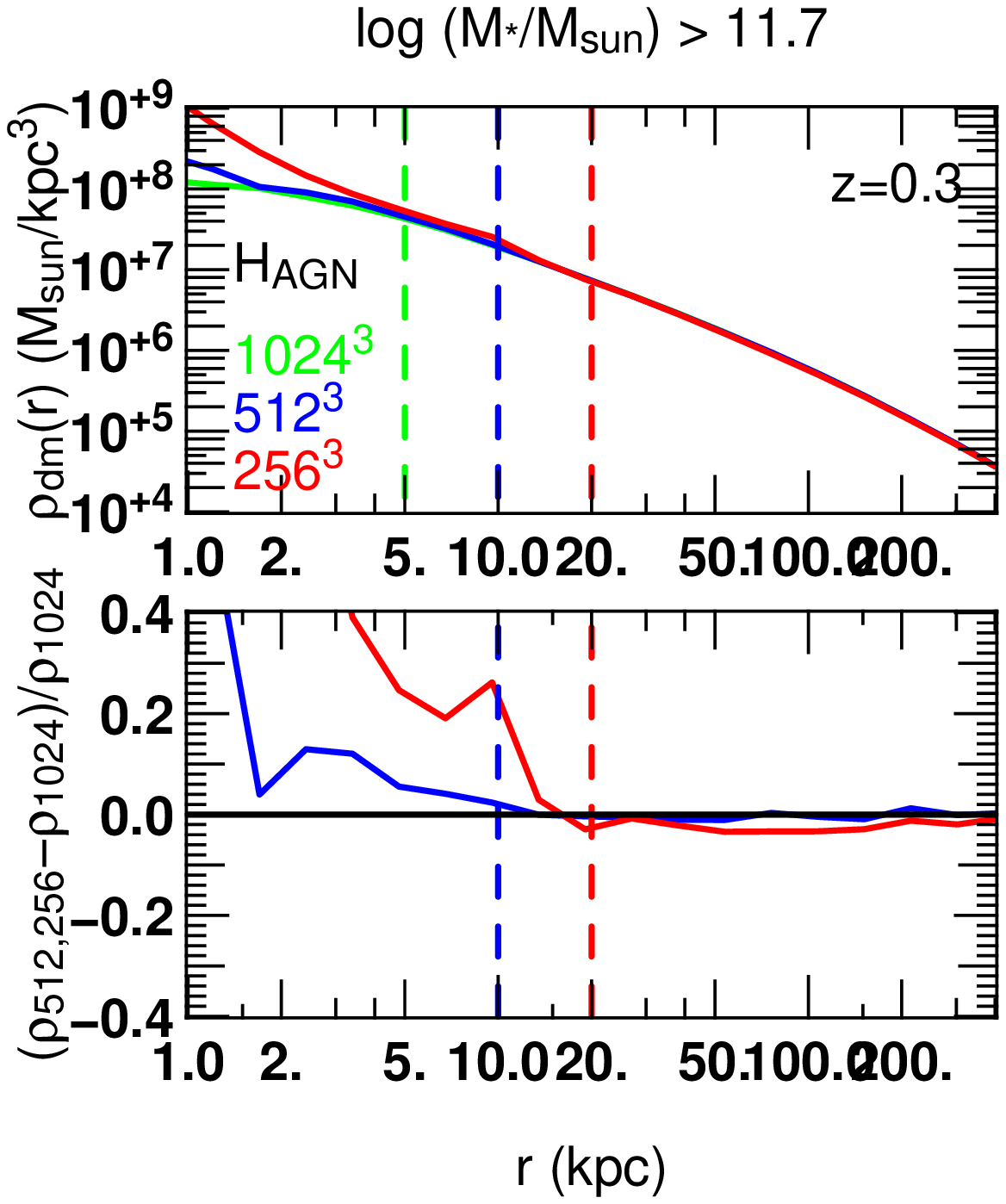}}
\caption{As in Fig. \ref{figapp1} but for DM haloes hosting a galaxy with a mass of
 $ 2\times 10^{11} M_\odot \geq M_* \geq 10^{11} M_\odot$ (left panel) or $M_* \geq 5\times 10^{11} M_\odot$ (right panel)
at $z=0.3$. Same trends are obtained.}
\label{figapp2}
\end{center}
 \end{figure*}

Fig \ref{figapp1} compares the mean dark matter density profile for dark matter haloes hosting 
galaxies with a mass $M_*$ defined by  $ 2\times 10^{11} M_\odot \geq M_* \geq 10^{11} M_\odot$ at $z=1$. In this case, we
have considered 474 matched objects. As we can see, resolution effects cause the density to be overestimated 
in the inner parts of dark matter haloes.
Moreover, as shown in Fig \ref{figapp2}, we get the same trends at  $z=0.3$ using a similar galaxy mass
interval or considering more massive galaxies ($M_* \geq 5\times 10^{11} M_\odot$). We have considered
here samples of 846 and 89 DM haloes respectively to derive the mean density profiles.

\cite{power++03} recommend to use lower limit values of $\sim$5 , 10 and 20 kpc 
respectively for \hagn, \hagnn (512$^3$) and \hagnn (256$^3$) and for
 our the studied halo/galaxy mass range.
Although their analysis concerns pure dark matter simulations only,
Figures \ref{figapp1} and \ref{figapp2} strongly suggest that lower limits defined by \cite{power++03}
give a clear and satisfactory indication of where the density profiles should converge.

\begin{figure*}
\begin{center}
\rotatebox{0}{\includegraphics[width=\columnwidth]{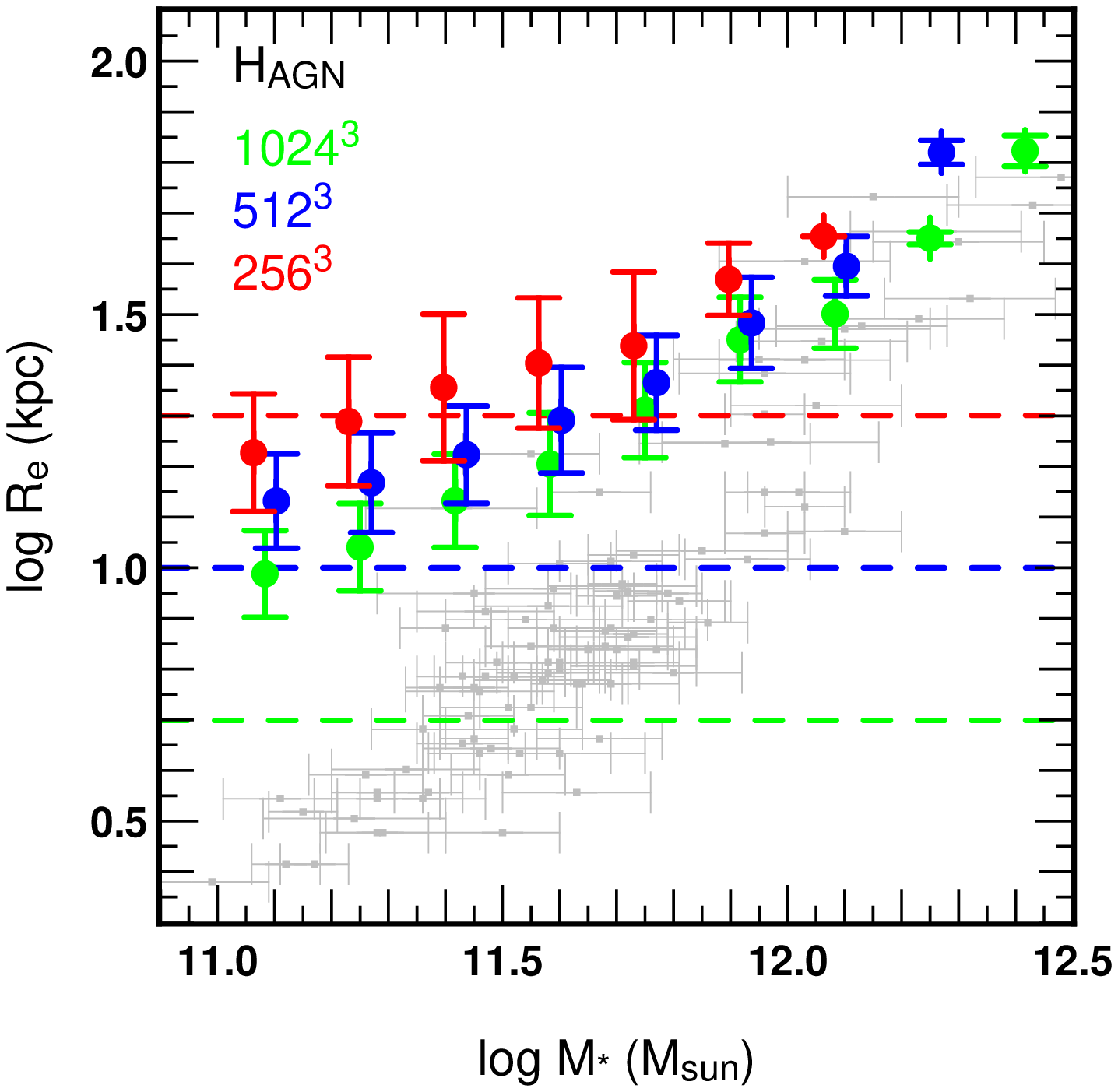}}
\rotatebox{0}{\includegraphics[width=\columnwidth]{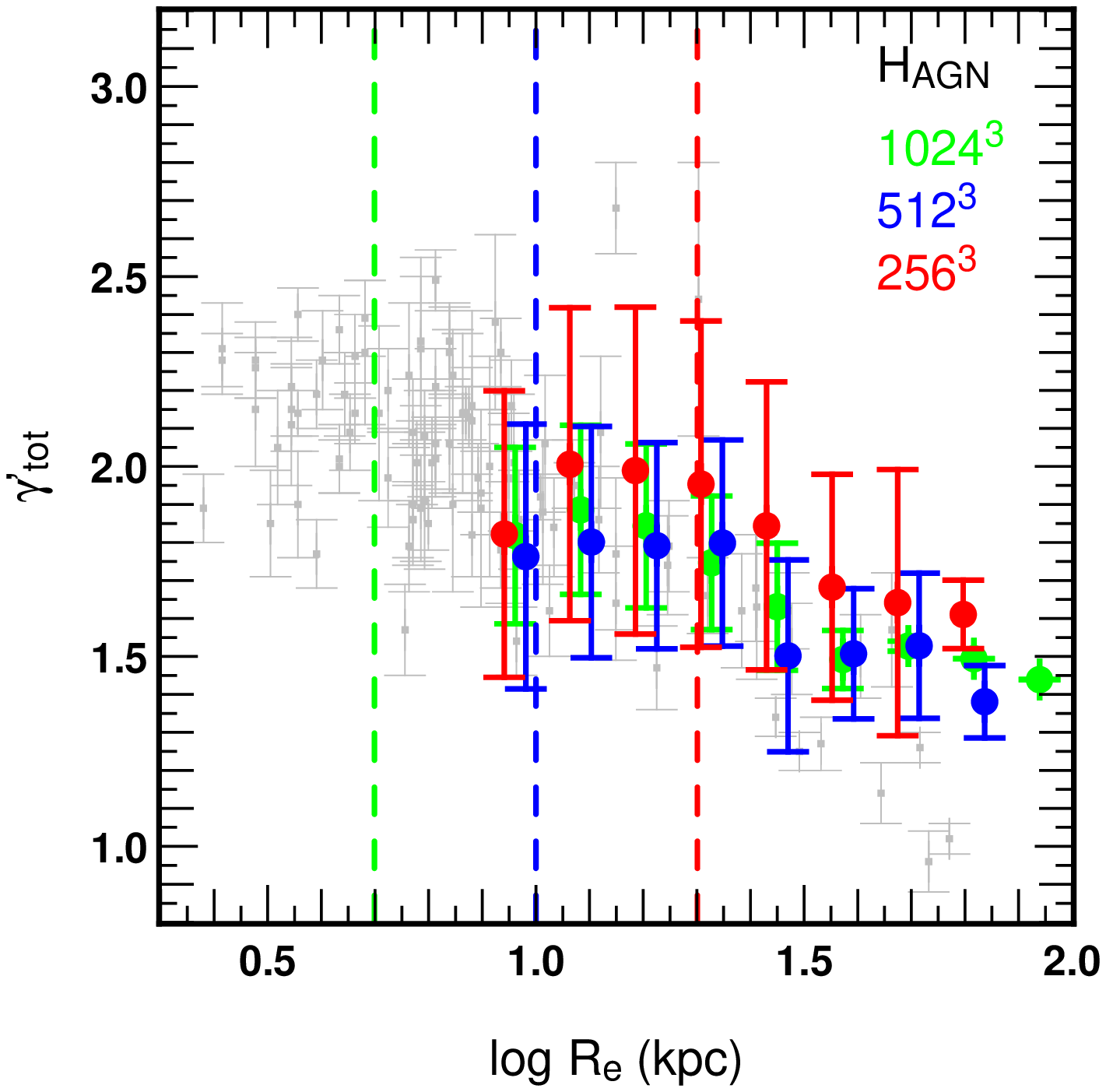}}
\rotatebox{0}{\includegraphics[width=7.6cm]{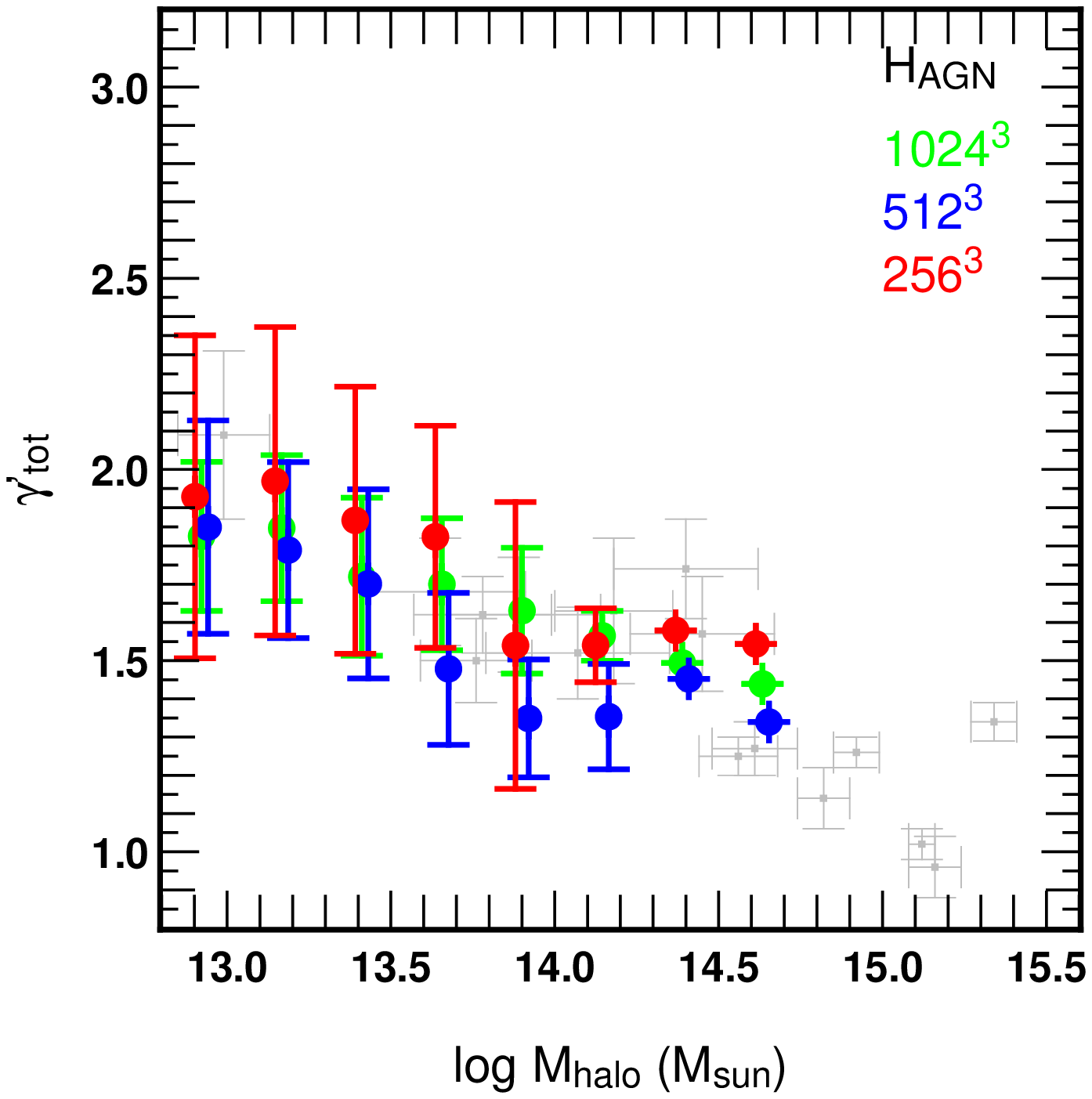}}
\rotatebox{0}{\includegraphics[width=\columnwidth]{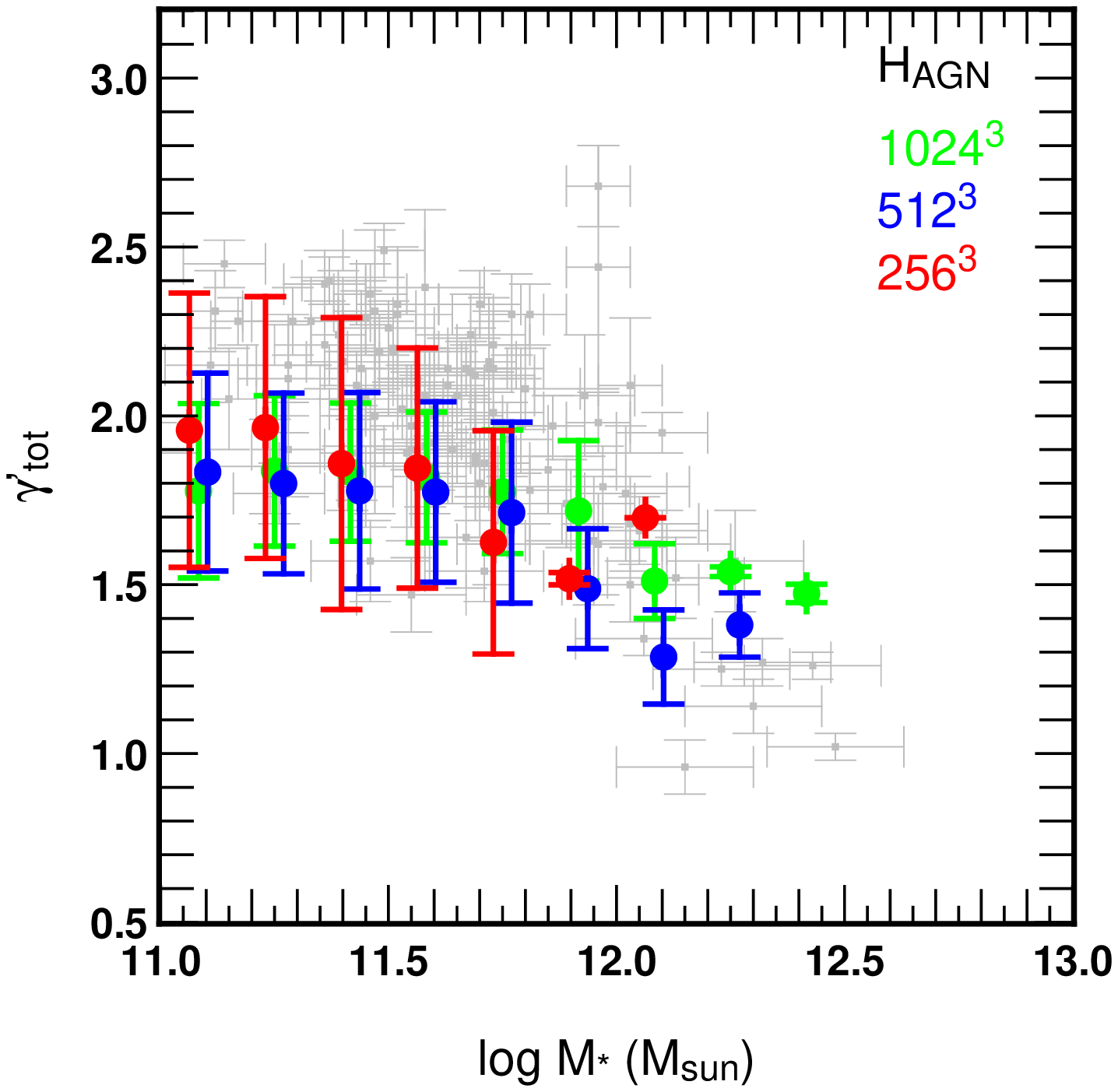}}
\caption{The variations of the effective radius $R_e$ with respect to the stellar mass $M_*$ (upper left panel) and
the variations of the mass-weighted total density slopes  $\gamma'_{tot}$ with respect
to the effective radius $R_{e}$ (upper right panel), to the dark matter halo masses $M_{halo}$ (lower left panel) and
to the stellar masses $M_*$ (lower right panel). Green color refers to our fiducial \hagnn simulation
while blue et red colors correspond to results from matching galaxies from \hagn-512$^3$  and
\hagn-256$^3$ respectively. The dashed lines (with same color code)  correspond 
to recommended lower limit values suggested by Power et al. (2003) for each simulation.
Grey data are observational data used in Fig. \ref{fig1}, \ref{fig2},  \ref{fig3} and \ref{fig4}. In spite of slight differences, same trends are obtained from the three runs in the considered galaxy mass range. 
}
\label{figapp3}
\end{center}
 \end{figure*}

Finally, we can also investigate the effect of resolution in the prediction of
observable quantities relevant to this work such as the effective radius $R_{e}$  and  the total density slope $\gamma'_{tot}$.
We therefore generate similar figures than Figs. \ref{fig1}, \ref{fig2},  \ref{fig3} and \ref{fig4}
with adding results derived from  \hagn-512$^3$ and  \hagn-256$^3$. One can see in Fig. \ref{figapp3} that 
the trends are quite similar between results from the three simulations, especially those relative to
the variations of $\gamma'_{tot}$ (upper right panel and lower ones).
However, there are some slight differences induced by the baryon component treatment. Indeed, higher resolutions 
imply in general higher gas densities especially in the central part of haloes which can lead to more 
gas cooling. Therefore, the galaxies produced in the higher resolution simulations
tend to be slightly more massive and more concentrated 
as the star formation tend to be more efficient in the inner regions. In this case,   such galaxies have on the average 
lower $R_{e}$ values as suggested by the upper left panel of Fig. \ref{figapp3}. However,
values of $R_{e}$ from the considered stellar mass range are quite close between the different simulations
and this seems not to particularly affect the estimation of $\gamma'$ in the interval [$R_{e}$/2 - $R_{e}$].

%%%%%%%%%%%%%%%%%%%%%%%%%%%%%%%%%%%%%%%%%%%%%%%%%
%     FIG appendix 1 : z=1 
%%%%%%%%%%%%%%%%%%%%%%%%%%%%%%%%%%%%%%%%%%%%%%%%%
%%%%%%%%%%%%%%%%%%%%%%%%%%%%%%%%%%%%%%%%%%%%%%%%%

%%%%%%%%%%%%%%%%%%%%%%%%%%%%%%%%%%%%%%%%%%%%%%%%%
%     FIG appendix 2 : z=0.3 logM~11 
%%%%%%%%%%%%%%%%%%%%%%%%%%%%%%%%%%%%%%%%%%%%%%%%%
%%%%%%%%%%%%%%%%%%%%%%%%%%%%%%%%%%%%%%%%%%%%%%%%%

%%%%%%%%%%%%%%%%%%%%%%%%%%%%%%%%%%%%%%%%%%%%%%%%%
%     FIG appendix 3 : convergence observations 
%%%%%%%%%%%%%%%%%%%%%%%%%%%%%%%%%%%%%%%%%%%%%%%%%
%%%%%%%%%%%%%%%%%%%%%%%%%%%%%%%%%%%%%%%%%%%%%%%%%

\end{document}